\documentclass[12pt,draftclsnofoot,journal,letterpaper,oneside,onecolumn]{IEEEtran}

\usepackage{amssymb}
\usepackage{amsmath}
\usepackage{amsfonts}
\usepackage{graphicx}
\usepackage{subfigure}
\usepackage{cite}
\usepackage{enumerate}
\usepackage{url}

\usepackage{algorithm}
\usepackage{algorithmic}

\usepackage[left=0.7in,right=0.7in,top=0.7in,bottom=0.7in]{geometry}

\begin{document}

\newtheorem{definition}{\bf Definition}
\newtheorem{proposition}{\bf Proposition}
\newtheorem{theorem}{\bf Theorem}
\newtheorem{lemma}{\bf Lemma}

\title{\LARGE{Distributed Cooperative Sensing in Cognitive Radio Networks: An Overlapping Coalition Formation Approach}}

\author{
\IEEEauthorblockN{
\small{Tianyu Wang}\IEEEauthorrefmark{1},
\small{Lingyang Song}\IEEEauthorrefmark{1},
\small{Zhu Han}\IEEEauthorrefmark{2},
\small{and Walid Saad}\IEEEauthorrefmark{3} \\}
\IEEEauthorblockA{
\IEEEauthorrefmark{1}\small{School of Electrical Engineering and Computer Science, Peking University, Beijing, China,} \\
\IEEEauthorrefmark{2}\small{Electrical and Computer Engineering Department, University of Houston, Houston, TX, USA,}\\
\IEEEauthorrefmark{3}\small{Electrical and Computer Engineering Department, University of Miami, Coral Gables, FL, USA,} \\}
}

\maketitle

\begin{abstract}

Cooperative spectrum sensing has been shown to yield a significant performance improvement in cognitive radio networks. In this paper, we consider distributed cooperative sensing (DCS) in which secondary users (SUs) exchange data with one another instead of reporting to a common fusion center. In most existing DCS algorithms, the SUs are grouped into disjoint cooperative groups or coalitions, and within each coalition the local sensing data is exchanged. However, these schemes do not account for the possibility that an SU can be involved in multiple cooperative coalitions thus forming \emph{overlapping} coalitions. Here, we address this problem using novel techniques from a class of cooperative games, known as \emph{overlapping coalition formation games}, and based on the game model, we propose a distributed DCS algorithm in which the SUs self-organize into a desirable network structure with overlapping coalitions. Simulation results show that the proposed overlapping algorithm yields significant performance improvements, decreasing the total error probability up to $25\%$ in the $Q_m+Q_f$ criterion, the missed detection probability up to $20\%$ in the $Q_m/Q_f$ criterion, the overhead up to $80\%$, and the total report number up to $10\%$, compared with the state-of-the-art non-overlapping algorithm.

\end{abstract}


%%%%%%%%%%%%%%%%%%%%%%%
\section{Introduction}%
%%%%%%%%%%%%%%%%%%%%%%%

Cognitive radio~(CR) has been proposed to increase spectrum efficiency, in which unlicensed, secondary users~(SUs), can sense the environment and change their parameters to access the spectrum of licensed, primary users~(PUs), while maintaining the interference to the PUs below a tolerable threshold~\cite{HNH-2009}. In order to exploit the spectrum holes, the SUs must be able to smartly sense the spectrum so as to decide which portion can be exploited~\cite{WL-2011}. Depending on the features of different signals, different spectrum sensing detectors have been designed, such as energy detectors, waveform-based detectors, and matched-filtering detectors~\cite{YA-2009}. However, the performance of these detectors is highly susceptible to the noise, small-scale fading, and shadowing over wireless channels. To overcome this problem, cooperative spectrum sensing~(CSS) was proposed, in which the SUs utilize the natural space diversity by sharing sensing results among each other and making collaborative decision on the detection of PUs~\cite{GS-2005, SZL-2007, PLGZ-2009, ZML-2009, FJS-2010, GL1-2007, GL2-2007, SHBDH-2011, WKJA-2010, GAK-2010, YZLXSG-2012, LXYZ-2013, L-2010, ZHLYP-2011, WGWY-2013, DWWSC-2013, MS-2008, CAP-2012, CAP-2013, CCPS-2013}. These existing works have shown that CSS can significantly improve the sensing accuracy, in comparison with the conventional, noncooperative case which relies solely on local detectors.

According to~\cite{ALB-2011}, the CSS schemes can be classified into three categories based on how the sensing data is shared in the network: centralized~\cite{GS-2005,SZL-2007,PLGZ-2009,ZML-2009,FJS-2010}, relay-assisted~\cite{GL1-2007,GL2-2007}, and distributed~\cite{SHBDH-2011, WKJA-2010, GAK-2010, YZLXSG-2012, LXYZ-2013}. In centralized CSS, a common fusion center (FC) collects sensing data from all the SUs in the network via a reporting channel, then combines the received local sensing data to determine the presence or absence of PUs, and at last diffuses the decision back to the SUs. In relay-assisted CSS, there is also a common FC, but the local sensing data, instead of being transmitted directly to the FC, is relayed by the SUs so as to reduce transmission errors. Unlike the centralized or relay-assisted CSS, distributed cooperative sensing (DCS) does not rely on an FC for making the cooperative decision. In this case, each SU simultaneously sends and receives sensing data via the reporting channel, and then combines the received data using a local fusion rule. Therefore, the SUs in DCS can make individual decisions on whether to access the spectrum, and thus, can adapt to the situation in which the SUs belong to different authorities or operators and distributed decisions must be made. Hereinafter, we focus on DCS.

In~\cite{SHBDH-2011}, the authors propose a coalition-based DCS, in which the SUs self-organize into \emph{disjoint} coalitions, and apply centralized CSS inside each coalition. The coalition formation process is based on a coalition formation game (CF-game) with nontransferable utility~\cite{SHDHB-2009, HNSBH-2011}, which jointly considers the associated benefit and cost for forming coalitions. This coalition-based DCS, in which the signaling overhead is shared by the coalition heads that are much closer to the SUs, can largely decrease the bandwidth requirement for reporting local sensing results. Other approaches that studied DCS are found in~\cite{WKJA-2010, GAK-2010, YZLXSG-2012, LXYZ-2013}. However, in~\cite{SHBDH-2011,WKJA-2010,GAK-2010,YZLXSG-2012,LXYZ-2013}, the network structure is \emph{restricted to disjoint, non-overlapping coalitions}, which implies that the local sensing results of an SU can only be shared within a single coalition, although, for the coalition-edge SUs, their local sensing results can be efficiently transmitted to the nearby coalitions for further improving the cooperative sensing performance. Hence, this disjoint coalitional structure of SUs may limit the gains from DCS and, thus, to reap the gains of DCS, information sharing among multiple coalition should be considered.

Traditionally, the SUs are assumed to share the same occupancy of PUs, i.e., whether the PU is present for all SUs or it is absent for all SUs. However, in practical systems, due to location and time diversities, the SUs may experience different spectrum occupancies. Some recent studies have noticed this problem and algorithms for spectrum-heterogeneous cognitive radio systems have been proposed~\cite{L-2010, ZHLYP-2011, WGWY-2013, DWWSC-2013}. Besides the diversity of SUs, other issues that greatly influence the sensing performance of SUs have also been studied, e.g., the spatial correlation between SUs~\cite{MS-2008, CAP-2012}, the mobility of PUs~\cite{CAP-2013}, the non-idealness of the report channel~\cite{CCPS-2013}. Although, these new features are not considered in this paper, we show that our model and analysis can be extended to involve such concerns by redefining the utility function.

The main contribution of this paper is to develop a novel DCS approach in which SUs can share their sensing information with a multitude of coalitions. In particular, we consider two criteria to evaluate the sensing performance, and for each criterion, we formulate the general DCS problem as an optimization with strict power and bandwidth constraints. In order to solve the DCS problem distributively, we introduce a new overlapping coalition formation (OCF) approach, which significantly differs from the existing non-overlapping DCS such as in ~\cite{SHBDH-2011} as it allows each SU to cooperate with \emph{multiple, overlapping coalitions} by allocating each coalition a portion of its local power and bandwidth resources. In particular, we introduce \emph{overlapping coalition formation games}~\cite{CEMJ-2010}, to model the DCS problem, and we propose a distributed algorithm that is shown to converge to a stable coalitional structure with overlapping coalitions. Simulation results show that the proposed overlapping algorithm yields significant performance improvements compared with the state-of-the-art non-overlapping algorithm for all network scenarios while also reducing the required overhead and system complexity.

The idea of using OCF-games to solve DCS problems is first introduced in our previous work~\cite{WSHS-2013}. In this paper, we consider a more practical system in which the power and bandwidth of each SU is limited, and we derive the OCF-game model directly from the formulated optimization problem, which guarantees that the coalition utility precisely represents the sensing performance of an SU. This fundamental difference makes the OCF-game model here more reasonable, and the corresponding algorithms and convergence analysis completely different from \cite{WSHS-2013}. In addition, we provide a separate section to discuss how some practical issues may effect our proposal and show how we can extend our model to involve such considerations, and extensive simulations are provided in various conditions.

The rest of this paper is organized as follows. In Section~\uppercase\expandafter{\romannumeral2}, we present the system model of generalized DCS with power and bandwidth constraints. In Section~\uppercase\expandafter{\romannumeral3}, the mathematical formulation of DCS is presented under two different criteria. In Sections~\uppercase\expandafter{\romannumeral4} and~\uppercase\expandafter{\romannumeral5}, we remodel the DCS problem via the OCF-game and CF-game, and propose two corresponding algorithms that converge to overlapping and non-overlapping coalitional structures, respectively. In Section~\uppercase\expandafter{\romannumeral6}, we consider some practical concerns. Simulation results of different algorithms under different criteria and constraints are compared and analyzed in Section~\uppercase\expandafter{\romannumeral7}. Finally, we draw our conclusions in Section~\uppercase\expandafter{\romannumeral8}.

%%%%%%%%%%%%%%%%%%%%%%%%%%%%%%%%%%%%%%%%%%%%%%%
\section{System Model}%
%%%%%%%%%%%%%%%%%%%%%%%%%%%%%%%%%%%%%%%%%%%%%%%

\begin{figure}[!t]
\centering
\includegraphics[width=2.6in]{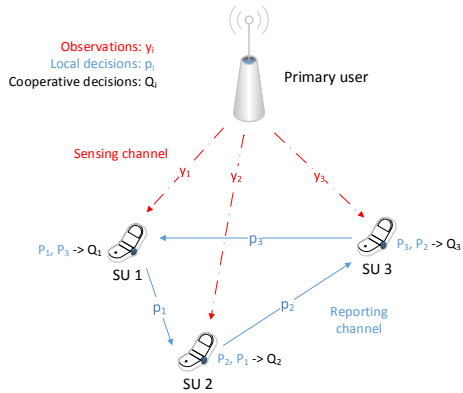}
\caption{Illustration of distributed cooperative sensing in a cognitive radio network with $3$ SUs.} \label{system_model}
\end{figure}

Consider a cognitive radio network with $N$ SUs equipped with energy detectors~\cite{PLGZ-2009}, the set of which is denoted by $\mathcal{N} = \{1,2,\ldots,N\}$, and a single PU far away from from them ~\cite{PLGZ-2009,ZML-2009}. The distance between SU $i$ and SU $j$ is denoted by $d_{i,j}$. The distance between the PU and any SU is denoted by $D$, and we have $D \gg d_{i,j}$ for any SU $i$ and $j$. In this network, the SUs individually and locally decide on the presence or absence of the PU via their own local information. We assume that the SUs can cooperate with one another by exchanging their sensing data via a reporting channel, and the overall DCS phase consists of three successive periods: the local sensing period, the data reporting period and the data fusion period. In the local sensing period, each SU locally detects the presence of the PU on the sensing channel. In the data reporting period, each SU sends its own sensing data to other SUs via the reporting channel with power and bandwidth constraints. In the data fusion period, each SU combines its local sensing data with the received sensing data and decides whether or not the PU is present. Once the DCS phase is completed, each SU locally decides whether to access the spectrum based on its decision of the PU' state as well as the particular distributed protocol used at the MAC layer, such as the distributed MAC protocols in \cite{ZTSC-2007,PZZ-2006}. Fig.~\ref{system_model} illustrates the DCS process described above in a CR network with $3$ SUs.

\subsection{Local Sensing}

We denote by $\mathcal{H}_1$ and $\mathcal{H}_0$ the hypotheses of the presence and absence of the PU, respectively. The sampled signal at SU $i\in \mathcal{N}$ is given by:
\begin{equation}
y_i(n) =
\begin{cases}
h_i(n) s(n) + u_i(n), & \mathcal{H}_1, \\
u_i(n), & \mathcal{H}_0,
\end{cases}
\end{equation}
where $h_i(n)$ denotes the channel between the PU and SU $i$, $s(n)$ denotes the signal from the PU and $u_i(n)$ denotes the noise at SU $i$.
In accordance with~\cite{PLGZ-2009}, we assume $s(n)$ is an independent identically distributed (i.i.d.) random process with zero mean and variance $\sigma_{s}^2$, $u_i(n)$ is i.i.d. Gaussian with zero mean and variance $\sigma_{u}^2$, $|h_i(n)|$ is Rayleigh distributed. Since the distances between any SUs are negligible compared with the distance from the PU to any SU, $|h_i(n)|, i \in \mathcal{N}$ are assumed to have the same variance $\sigma_{h}^2 = \kappa D^{-\mu}$, where $\kappa$ and $\mu$ are path loss parameters. For any SU $i \in \mathcal{N}$, the energy detector's probabilities of missed detection and false alarm are, respectively, given by~\cite{PLGZ-2009}:
\begin{equation} \label{Pm}
P_{m,i}(\lambda_i) = 1 - \mathcal{Q}\left( \left(\frac{\lambda_i}{1+\gamma} - 1\right) \sqrt{N_s} \right),
\end{equation}
\begin{equation} \label{Pf}
P_{f,i}(\lambda_i) = \mathcal{Q}\left( \left(\lambda_i - 1\right) \sqrt{N_s} \right),
\end{equation}
where $\mathcal{Q}(\cdot)$ denotes the right-tail probability of a normalized Gaussian distribution, $\gamma = \sigma_{h}^2 \sigma_{s}^2 / \sigma_{u}^2$ is the average received SNR at each SU, $\lambda_i \sigma_{u}^2$ is the threshold of the energy detector at SU $i$, and $N_s$ is the product of the sensing time and sampling frequency. We assume $\gamma$ and $N_s$ are constant parameters.

\subsection{Data Reporting}

In order to reduce the bandwidth for reporting, the local sensing data is quantized to $1$ bit (hard decisions) in~\cite{SZL-2007}. In addition, we assume that each SU has limited transmit power $P_{SU}$ and time-frequency resource $\theta_{SU}$ during the data reporting period. For the reporting between any two SUs, the minimum average received SNR is assumed to be $\gamma_0$ and the minimum time-frequency resource for transmitting $1$ bit is assumed to be $\theta_0$. Therefore, the power and bandwidth constraints for any SU $i \in \mathcal{N}$ are given by:
\begin{equation}
\sum\limits_{j \in \mathcal{S}_i}{ \frac{\gamma_0 \sigma_{u}^2}{\kappa d_{i,j}^{-\mu}} } \le P_{SU},
\end{equation}
\begin{equation}
|\mathcal{S}_i| \theta_0 \le \theta_{SU},
\end{equation}
where $\mathcal{S}_i$ is the ``report-to" set of SU $i$ consisting of the SUs that SU $i$ reports to.

\subsection{Data Fusion}

After every SU sends the sensing data to its designated receivers, each SU combines all the received sensing data (including its local sensing data) using a local fusion rule. Suppose that the $k_i$-out-of-all fusion rule is adopted by SU $i$ \cite{PLGZ-2009,ZML-2009}, i.e., SU $i$ decides the presence of the PU if at least $k_i$ reports declare that the PU is detected, and vice versa. Consequently, SU $i$'s probabilities of missed detection and false alarm are, respectively, given by:
\begin{equation} \label{Qm}
Q_{m,i}(k_i) = \sum\limits_{|\mathcal{R}_i^1|<k_i}{ \left[\prod\limits_{j\in \mathcal{R}_i^1}{(1-P_{m,j})}\prod\limits_{j\in \mathcal{R}_i^0}{P_{m,j}}\right]},
\end{equation}
\begin{equation} \label{Qf}
Q_{f,i}(k_i) = \sum\limits_{|\mathcal{R}_i^1|\ge k_i}{ \left[\prod\limits_{j\in \mathcal{R}_i^1}{P_{f,j}}\prod\limits_{j\in \mathcal{R}_i^0}{(1-P_{f,j})}\right]},
\end{equation}
where $\mathcal{R}_i^1 \cup \mathcal{R}_i^0 = \mathcal{R}_i$ is the ``report-from" set of SU $i$ which consists of SU $i$ it as well as the SUs that report to SU $i$, and $\mathcal{R}_i^1, \mathcal{R}_i^0$ denote the set of SUs whose reports declare the presence and absence of the PU, respectively. Note that $k_i$ is an integer between $1$ and $|\mathcal{R}_i|$.

%%%%%%%%%%%%%%%%%%%%%%%%%%%%%%%%%%%%%%%%%%%%%%%%%%%%%%%%%%%%%%%%%%%%%%%%%%%%%%%%%%%%%%%%%%%%%%%%%%%%%%%%%%%%%%%%
\section{DCS as an Optimization Problem}%
%%%%%%%%%%%%%%%%%%%%%%%%%%%%%%%%%%%%%%%%%%%%%%%%%%%%%%%%%%%%%%%%%%%%%%%%%%%%%%%%%%%%%%%%%%%%%%%%%%%%%%%%%%%%%%%%

From the system model in Section~\uppercase\expandafter{\romannumeral2}, we can see that the DCS process is determined by local parameters as well as by the reporting structure of the network, i.e., the local sensing thresholds $\lambda_i, i \in \mathcal{N}$, the local fusion rules $k_i, i \in \mathcal{N}$, and the report-to sets $\mathcal{S}_i, i \in \mathcal{N}$, or equally, the report-from sets $\mathcal{R}_i, i \in \mathcal{N}$. We consider a $1\times N$ vector $\mathbf{\Lambda} = (\lambda_1,\lambda_2,\ldots,\lambda_N)$ as the local threshold vector, a $1\times N$ vector $\mathbf{K} = (k_1,k_2,\ldots,k_N)$ as the fusion rule vector, and an $N \times N$ binary matrix $\mathbf{\Omega} = \{\omega_{i,j}\}, \omega_{i,j} = \{0,1\}$ as the reporting matrix where $\omega_{i,j} = 1$ implies that SU $i$'s report is received by SU $j$. Note that the ``report-to" sets $\mathcal{S}_i,i \in \mathcal{N}$ and the ``report-from" sets $\mathcal{R}_i,i \in \mathcal{N}$ are given by the rows and columns of $\mathbf{\Omega}$, respectively. To evaluate the performance of DCS, we consider two criteria that are commonly used in the literature, the $Q_m+Q_f$ criterion~\cite{ZML-2009} and $Q_m/Q_f$ criterion~\cite{SHBDH-2011}.

\subsection{$Q_m+Q_f$ Criterion}

In the $Q_m+Q_f$ criterion, we consider the probability that the cooperative sensing decision is incorrect, which is referred to as the ``total error rate" in~\cite{ZML-2009}. Strictly speaking, the total error rate of an SU $i \in \mathcal{N}$ is given by $P_1 Q_{m,i} + (1-P_1) Q_{f,i} $, where $P_1$ is the probability that the PU is present. For conciseness, we assume $P_1 =0.5$, and thus, the total error rate is given by $(Q_{m,i} + Q_{f,i})/2$. Moreover, we consider the average sensing performance of all SUs in the network, i.e., $(1/2N) \sum\nolimits_{i \in \mathcal{N}} (Q_{m,i} + Q_{f,i})$. By omitting the factor $1/2N$ from the objective function, we have the DCS problem is formulated as:
\begin{subequations} \label{Opt1}
\begin{equation} \label{Opt1a}
\min\limits_{\atop \mathbf{\Lambda},\mathbf{K},\mathbf{\Omega}} \sum\limits_{i\in \mathcal{N}}{\left(Q_{m,i} + Q_{f,i}\right)},
\end{equation}
\begin{align}
s.t. &~ \sum\limits_{j\ne i \atop \omega_{i,j} = 1} \frac{\gamma_0 \sigma_u^2}{\kappa d_{i,j}^{-\mu}} \le P_{SU}, ~i = 1,2,\ldots,N, \label{Opt1b}\\
     &~ \sum\limits_{j\ne i \atop \omega_{i,j} = 1} \theta_0 \le \theta_{SU}, ~i = 1,2,\ldots,N, \label{Opt1c}
\end{align}
\end{subequations}
where $Q_{m,i}$ and $Q_{f,i}$ are given by (\ref{Qm}) and (\ref{Qf}), respectively. We note that our model and analysis can be extended to the more general setting with any $P_1$, in a straightforward manner, and our results still hold.

Problem (\ref{Opt1}) is a mixed integer nonlinear programming problem that is known to be intractable in the general case~\cite{HW-2006}. Moreover, due to the lack of a fusion or control center in the considered network, any possible centralized algorithm that gives an optimal solution will not be applicable for the DCS process. Therefore, we consider suboptimal solutions with distributed algorithms. Note that the constraints (\ref{Opt1b}) and (\ref{Opt1c}) are only related to $\mathbf{\Omega}$; we consider a suboptimal solution with two separate steps:
\begin{enumerate} [a)]
    \item Find a feasible reporting matrix $\mathbf{\Omega}$ that satisfies the constraints in (\ref{Opt1b}) and (\ref{Opt1c}).
    \item Compute the optimal $\mathbf{\Lambda}$ and $\mathbf{K}$ for the objective function (\ref{Opt1a}) with $\mathbf{\Omega}$ given in step a.
\end{enumerate}

To simplify the problem, we assume that the AND rule is adopted by all SUs, i.e., $k_i = |\mathcal{R}_i|, i \in \mathcal{N}$, and thus, step b is reduced to the computation of the optimal $\mathbf{\Lambda}$ for the objective function with the given $\mathbf{\Omega}$ and ${\mathbf{K}} = (|\mathcal{R}_1|, |\mathcal{R}_2|, \ldots, |\mathcal{R}_N|)$ where $|\mathcal{R}_i| = \sum\nolimits_{j\in \mathcal{N}}{\omega_{j,i}}, i \in \mathcal{N}$. By substituting $\mathbf{\Omega}, \mathbf{K}$ and (\ref{Pm}), (\ref{Pf}) into (\ref{Qm}), (\ref{Qf}), and further substituting (\ref{Qm}) and (\ref{Qf}) into (\ref{Opt1a}), step b is formally written as:
\begin{equation} \label{Opt12}
\min\limits_{\atop \mathbf{\Lambda}} \sum\limits_{i\in \mathcal{N}}{ \left[1- \prod\limits_{j \in \mathcal{N} \atop \omega_{j,i} =1 }\mathcal{Q}\left( \left(\frac{\lambda_j}{1+\gamma} - 1\right) \sqrt{N_s} \right) + \prod\limits_{j \in \mathcal{N} \atop \omega_{j,i} =1 }\mathcal{Q}\left( \left(\lambda_j - 1\right) \sqrt{N_s} \right) \right] }.
\end{equation}

\subsection{$Q_m/Q_f$ Criterion}

In the $Q_m/Q_f$ criterion, the network sensing performance is evaluated via the average value of one error probability while the other probability is maintained below a certain threshold $\alpha$. In this paper, we consider the average value of the missed detection probability while the false alarm probabilities are such that $Q_{f,i} \le \alpha, i \in \mathcal{N}$. This criterion indicates the interference to the PU while we guarantee a usability rate of the spectrum holes. Mathematically, the DCS problem is formulated as:
\begin{subequations} \label{Opt2}
\begin{equation} \label{Opt2a}
\min\limits_{\atop \mathbf{\Lambda},\mathbf{K},\mathbf{\Omega}} \sum\limits_{i\in \mathcal{N}}{Q_{m,i}},
\end{equation}
\begin{align}
s.t. &~ \sum\limits_{j\ne i \atop \omega_{i,j} = 1} \frac{\gamma_0 \sigma_u^2}{\kappa d_{i,j}^{-\mu}} \le P_{SU}, ~i = 1,2,\ldots,N, \label{Opt2b}\\
     &~ \sum\limits_{j\ne i \atop \omega_{i,j} = 1} \theta_0 \le \theta_{SU}, ~i = 1,2,\ldots,N, \label{Opt2c} \\
     &~ Q_{f,i} \le \alpha, ~i = 1,2,\ldots,N, \label{Opt2d}
\end{align}
\end{subequations}
where $Q_{m,i}$ and $Q_{f,i}$ are given by (\ref{Qm}) and (\ref{Qf}), respectively.

Problem (\ref{Opt2}) is also a mixed integer nonlinear programming problem. For similar reasons as in the $Q_m+Q_f$ criterion, we consider a suboptimal solution with two separate steps:
\begin{enumerate} [a)]
    \item Find a feasible reporting matrix $\mathbf{\Omega}$ that satisfies the constraints in in (\ref{Opt2b}) and (\ref{Opt2c}).
    \item Compute the optimal $\mathbf{\Lambda}$, $\mathbf{K}$ for (\ref{Opt2a}) with $\mathbf{\Omega}$ given in step a and the constrains in (\ref{Opt2d}).
\end{enumerate}

Note that step a is exactly the same as in the $Q_m+Q_f$ criterion. For step b, we also assume the AND rule is adopted by all SUs and , thus, it reduces to:
\begin{subequations} \label{Opt22}
\begin{align}
\min\limits_{\atop \mathbf{\Lambda}} &~ \sum\limits_{i\in \mathcal{N}}{ \left[1- \prod\limits_{j \in \mathcal{N} \atop \omega_{j,i} =1 }\mathcal{Q}\left( \left(\frac{\lambda_j}{1+\gamma} - 1\right) \sqrt{N_s} \right) \right] } \label{Opt22a} \\
s.t. &~ \prod\limits_{j \in \mathcal{N} \atop \omega_{j,i} =1 }\mathcal{Q}\left( \left(\lambda_j - 1\right) \sqrt{N_s} \right) \le \alpha, ~i = 1,2,\ldots,N  \label{Opt22b}.
\end{align}
\end{subequations}

%%%%%%%%%%%%%%%%%%%%%%%%%%%%%%%%%%%%%%%%%%%%%%%%%%%%%%%%%%%%%%%%%%%%%%%%%%%%%%%%%%%%%%%%%%%%%%%%%%%%%%%%%%%%%%%%
\section{DCS based on Overlapping Coalition Formation Games}%
%%%%%%%%%%%%%%%%%%%%%%%%%%%%%%%%%%%%%%%%%%%%%%%%%%%%%%%%%%%%%%%%%%%%%%%%%%%%%%%%%%%%%%%%%%%%%%%%%%%%%%%%%%%%%%%%

In the previous section, the DCS problem is divided into two separate subproblems. The first subproblem aims to find a feasible reporting matrix $\mathbf{\Omega}$. The second subproblem aims at computing the optimal sensing threshold vector $\mathbf{\Lambda}$ with the given $\mathbf{\Omega}$. In this section, we consider the DCS problem as an OCF-game, in which the first subproblem is strictly modeled as the local resource limitation, and the second subproblem is captured by an adequately designed utility function for each coalition of SUs. Based on the proposed OCF-game model, we propose a distributed coalition formation algorithm that allows to form overlapping coalitions, and a threshold decision algorithm that locally decides the sensing threshold of each SU. Note that the DCS problems in both the $Q_m+Q_f$ and $Q_m/Q_f$ criteria are uniformly modeled by the OCF-game. The proposed algorithms apply to both criteria.

\subsection{OCF-game Model}

In essence, coalitional games involve a set of players who seek to form cooperative groups, i.e., coalitions, to strengthen their positions in a given game scenario~\cite{SHDHB-2009}. In particular, in an OCF-game~\cite{CEMJ-2010}, the players can join multiple coalitions by contributing parts of their limited resources to different coalitions. Each coalition constitutes a group of players who are working together and whose utility is captured by both a coalition-level value and an individual user payoff. In a coalition formation game, each player individually decides which coalitions it wishes to join, so as to maximize its total payoff with the limited resources. Note that the coalitions can, in general, be overlapping, such that a player can participate in multiple coalitions simultaneously.

For the DCS problem, the players are the SUs $\mathcal{N} = \{1,2,\ldots,N\}$ with power resource $P_{SU}$ and bandwidth resource $\theta_{SU}$. Here, a coalition $\mathcal{R}_i \subseteq \mathcal{N}$ denotes a cooperative group of SUs in which the coalition members report their sensing results to a given SU $i\in \mathcal{N}, i \in \mathcal{R}_i$. The power and bandwidth resources contributed by player $j \ne i, j\in \mathcal{R}_i$ to coalition $\mathcal{R}_i$ are $(\gamma_0 \sigma_{u}^2) / (\kappa d_{i,j}^{-\mu})$ and $\theta_0$, respectively. Also, player $i \in \mathcal{N}$ naturally belongs to coalition $\mathcal{R}_i$ without contributing any power or bandwidth resource. Note that coalitions $\mathcal{R}_i$ and $\mathcal{R}_j, j \ne i$ can be exactly the same when SUs $i$ and $j$ receive the sensing results from the same SUs. However, we still treat them as two different coalitions and differentiate them with different subscripts, because:
\begin{enumerate} [(1)]
    \item Coalitions $\mathcal{R}_i$ and $\mathcal{R}_j$ represent the received sensing results at different SUs, i.e., SU $i$ and SU $j$.
    \item For any SU $k$ belonging to both coalitions, coalitions $\mathcal{R}_i$ and $\mathcal{R}_j$ require different power resource contributions, i.e., $(\gamma_0 \sigma_{u}^2) / (\kappa d_{i,k}^{-\mu})$ and $(\gamma_0 \sigma_{u}^2) / (\kappa d_{j,k}^{-\mu})$.
\end{enumerate}
For all $N$ SUs in the network, there are exactly $N$ coalitions that correspond to them. For the completeness of the game model, we define a \emph{coalitional structure} as the set of all coalitions, denoted by $\mathcal{CS} = \{\mathcal{R}_1,\mathcal{R}_2,\ldots,\mathcal{R}_N\}$. Note that $\mathcal{CS}$ is just another expression of the reporting matrix $\mathbf{\Omega}$.

\begin{figure}
\centering
\subfigure[the $Q_m+Q_f$ criterion]{
\label{utility_a} %% label for first subfigure
\includegraphics[width=2.5in]{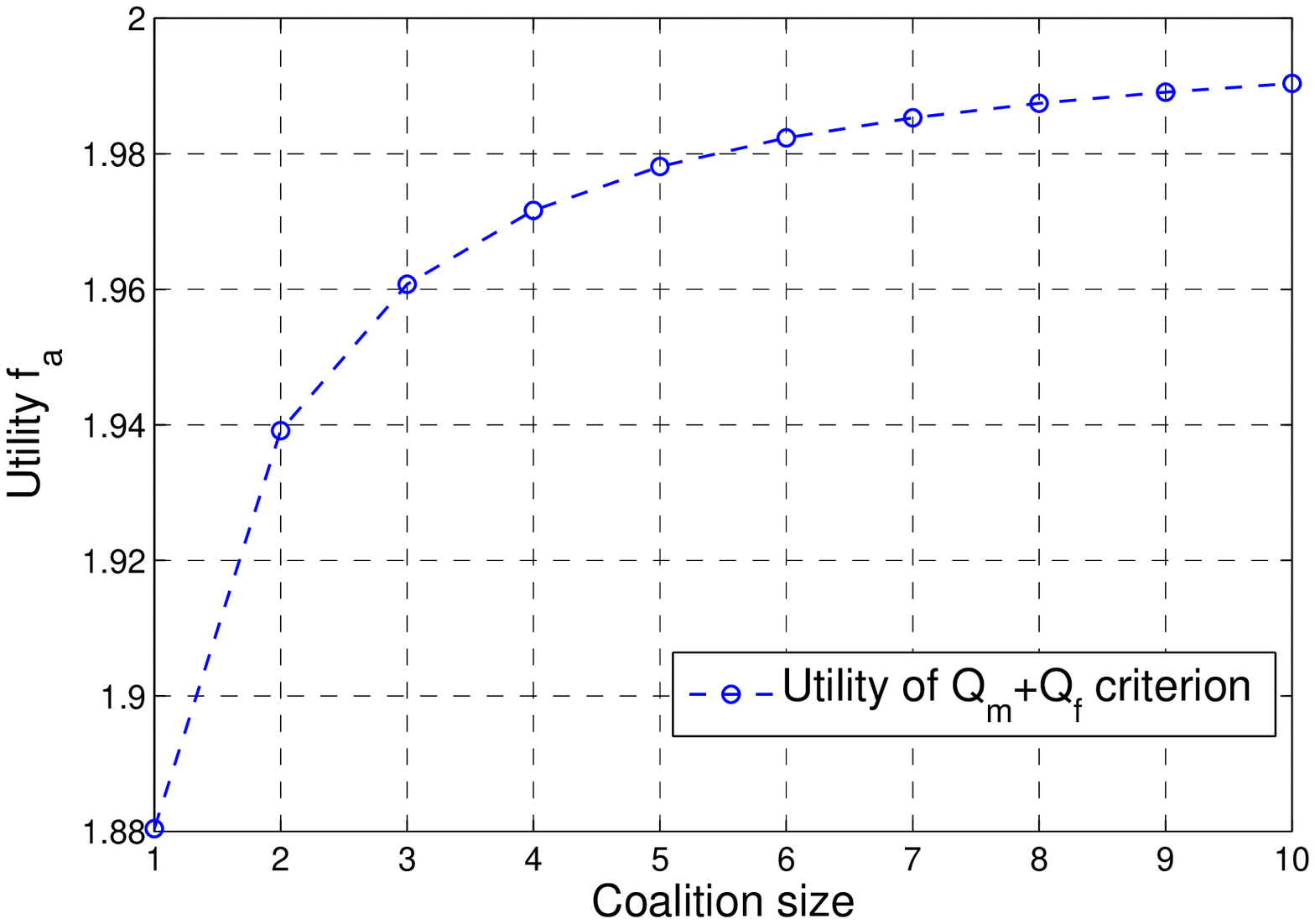}}
\hspace{-0.3in}
\subfigure[the $Q_m/Q_f$ criterion]{
\label{utility_b} %% label for second subfigure
\includegraphics[width=2.5in]{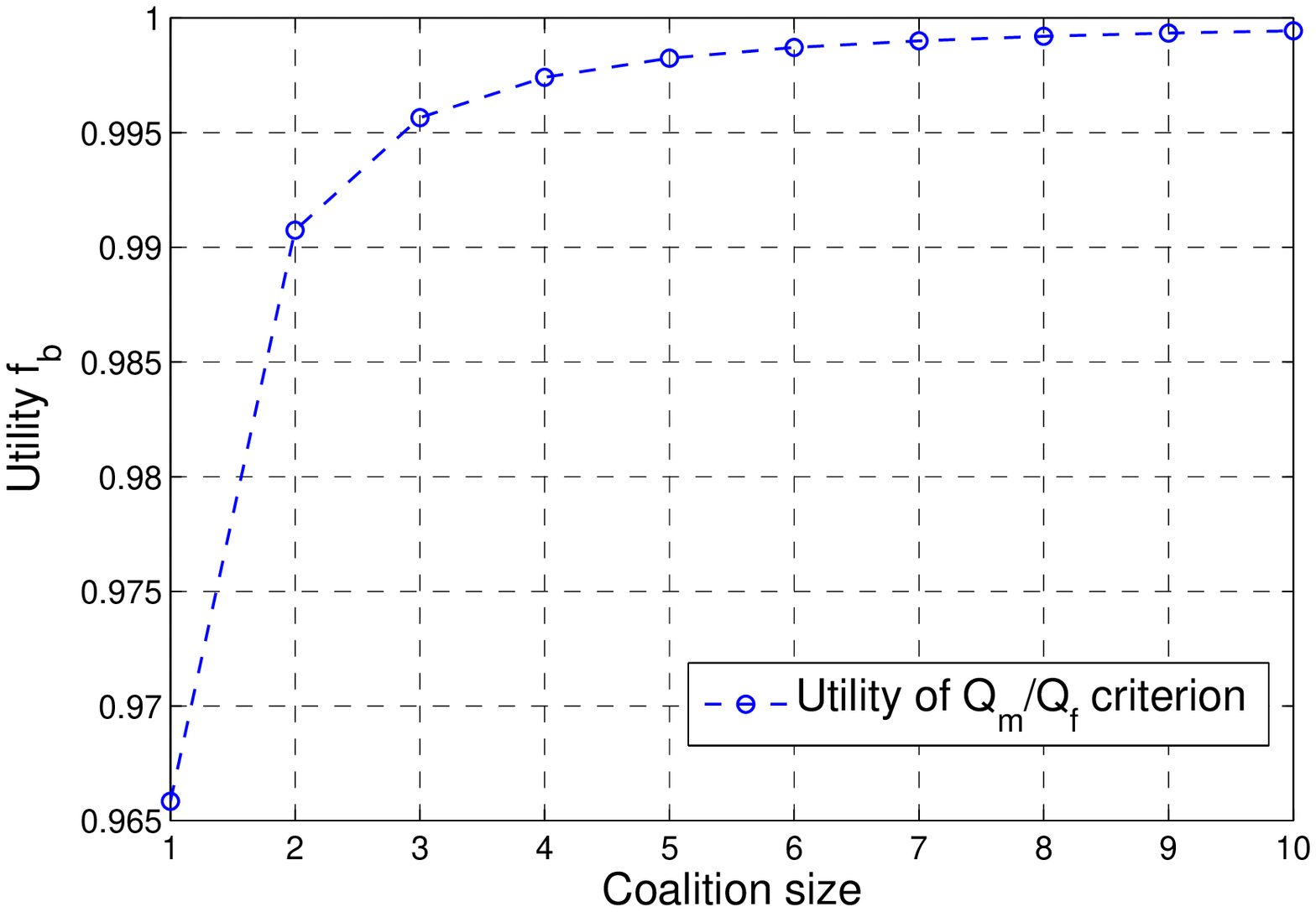}}
\caption{Coalition utility as a function of coalition size for both the $Q_m+Q_f$ and $Q_m/Q_f$ criteria.}
\label{utility_ab} %% label for entire figure
\end{figure}

To capture the performance of a given coalition $\mathcal{R}_i$, we propose a utility function that captures the best sensing performance of SU $i$, given by:
\begin{equation} \label{utilityOrigin}
U(\mathcal{R}_i) =
\begin{cases}
2 -  \min\limits_{\atop \mathbf{\Lambda}(\mathcal{R}_i)} {\left(Q_{m,i} + Q_{f,i}\right)}, &~ \text{$Q_m+Q_f$ criterion}, \\
1 -  \min\limits_{\atop \mathbf{\Lambda}(\mathcal{R}_i), Q_{f,i} \le \alpha } {Q_{m,i}}, &~ \text{$Q_m/Q_f$ criterion},
\end{cases}
\end{equation}
where $Q_{m,i}$ and $Q_{f,i}$ are given by (\ref{Qm}) and (\ref{Qf}) with $k_i = |\mathcal{R}_i|$, and $\mathbf{\Lambda}(\mathcal{R}_i)$ is the local sensing threshold vector for the players in $\mathcal{R}_i$. Note that we use ``$2-$" and ``$1-$" to maintain the utility to be positive, since all the probabilities $Q_{m,i}$ and $Q_{f,i}$ are between $0$ and $1$. Due to the symmetry of $U(\mathcal{R}_i)$ to the members in $\mathcal{R}_i$, we point out that the optimal value is obtained when all the coalition members have the same local sensing threshold, i.e., $\mathbf{\Lambda}(\mathcal{R}_i) = (\lambda, \lambda, \ldots, \lambda)_{1\times |\mathcal{R}_i|}$. For the $Q_m+Q_f$ criterion, the optimal threshold $\lambda_{a}(|\mathcal{R}_i|)$ and the value of (\ref{utilityOrigin}) $f_a(|\mathcal{R}_i|)$ are given in Appendix A. For the $Q_m/Q_f$ criterion, the optimal threshold $\lambda_{b}(|\mathcal{R}_i|)$ and the value of (\ref{utilityOrigin}) $f_b(|\mathcal{R}_i|)$ are given in Appendix B. Then, we have
\begin{equation} \label{optLambda}
\mathbf{\Lambda}(\mathcal{R}_i) =
\begin{cases}
(\lambda_{a}(|\mathcal{R}_i|), \lambda_{a}(|\mathcal{R}_i|), \ldots, \lambda_{a}(|\mathcal{R}_i|))_{1\times |\mathcal{R}_i|}, &~ \text{$Q_m+Q_f$ criterion}, \\
(\lambda_{b}(|\mathcal{R}_i|), \lambda_{b}(|\mathcal{R}_i|), \ldots, \lambda_{b}(|\mathcal{R}_i|))_{1\times |\mathcal{R}_i|}, &~ \text{$Q_m/Q_f$ criterion},
\end{cases}
\end{equation}
and
\begin{equation} \label{utility}
U(\mathcal{R}_i) = U(|\mathcal{R}_i|) =
\begin{cases}
f_a(|\mathcal{R}_i|), &~ \text{$Q_m+Q_f$ criterion}, \\
f_b(|\mathcal{R}_i|), &~ \text{$Q_m/Q_f$ criterion}.
\end{cases}
\end{equation}
Note that $U(\mathcal{R}_i)$ is only determined by the coalition size, and its value is limited and discrete. The numerical results in Fig.~\ref{utility_ab} show that $U(|\mathcal{R}_i|)$ is an increasing concave function in both criteria, i.e.,
\begin{equation} \label{property1}
U(|\mathcal{R}_i|) > U(|\mathcal{R}_j|), \text{ with } |\mathcal{R}_i|>|\mathcal{R}_j|,
\end{equation}
and
\begin{equation} \label{property2}
U(|\mathcal{R}_i|) - U(|\mathcal{R}_i|-1) < U(|\mathcal{R}_j|) - U(|\mathcal{R}_j|-1), \text{ with } |\mathcal{R}_i|>|\mathcal{R}_j|.
\end{equation}

The utility function (\ref{utilityOrigin}) captures the sensing performance of SU $i$ when all members in $\mathcal{R}_i$ report to SU $i$ by using the corresponding power and bandwidth resources. The network sensing performance, which is the average value of the SUs' sensing performance, therefore, is captured by the social welfare, defined as the sum utility of all the coalitions, given by:
\begin{equation}
\Upsilon(\mathcal{CS}) = \sum\limits_{\mathcal{R}_i \in \mathcal{CS}}U(\mathcal{R}_i).
\end{equation}
Considering the monotone-increasing property of $U(\cdot)$, as given by (\ref{property1}), we can expect a larger social welfare, or equally, a better network sensing performance, as the average coalition size increases. However, the power cost for a SU joining a coalition increases with the distance between the SU and the coalition. Thus, due to the limited power of each SU, the \emph{grand coalition} that includes all SUs seldom forms.

\begin{figure}
\centering
\subfigure[the $Q_m+Q_f$ criterion]{
\label{phi_a} %% label for first subfigure
\includegraphics[width=2.5in]{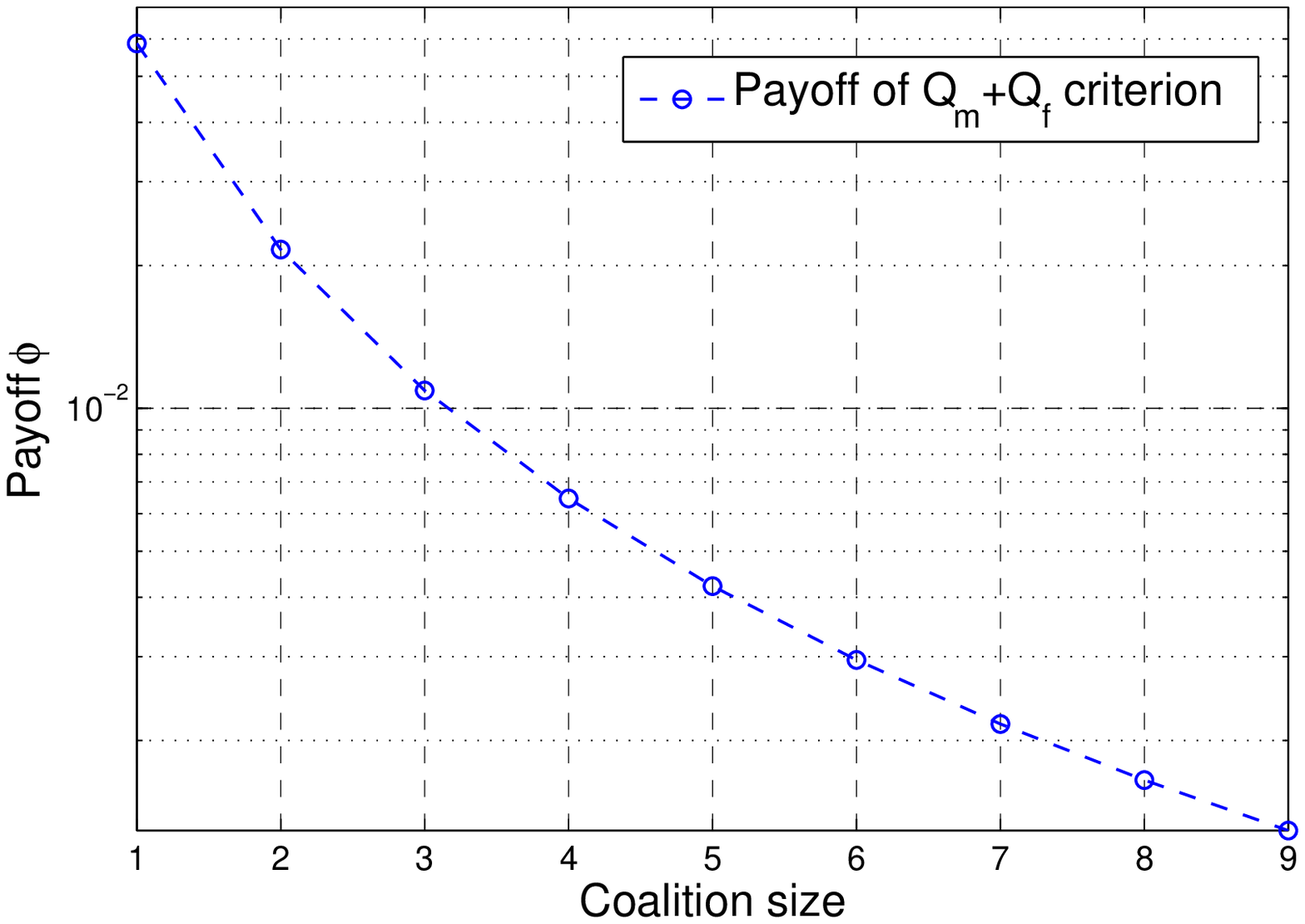}}
\hspace{-0.3in}
\subfigure[the $Q_m/Q_f$ criterion]{
\label{phi_b} %% label for second subfigure
\includegraphics[width=2.5in]{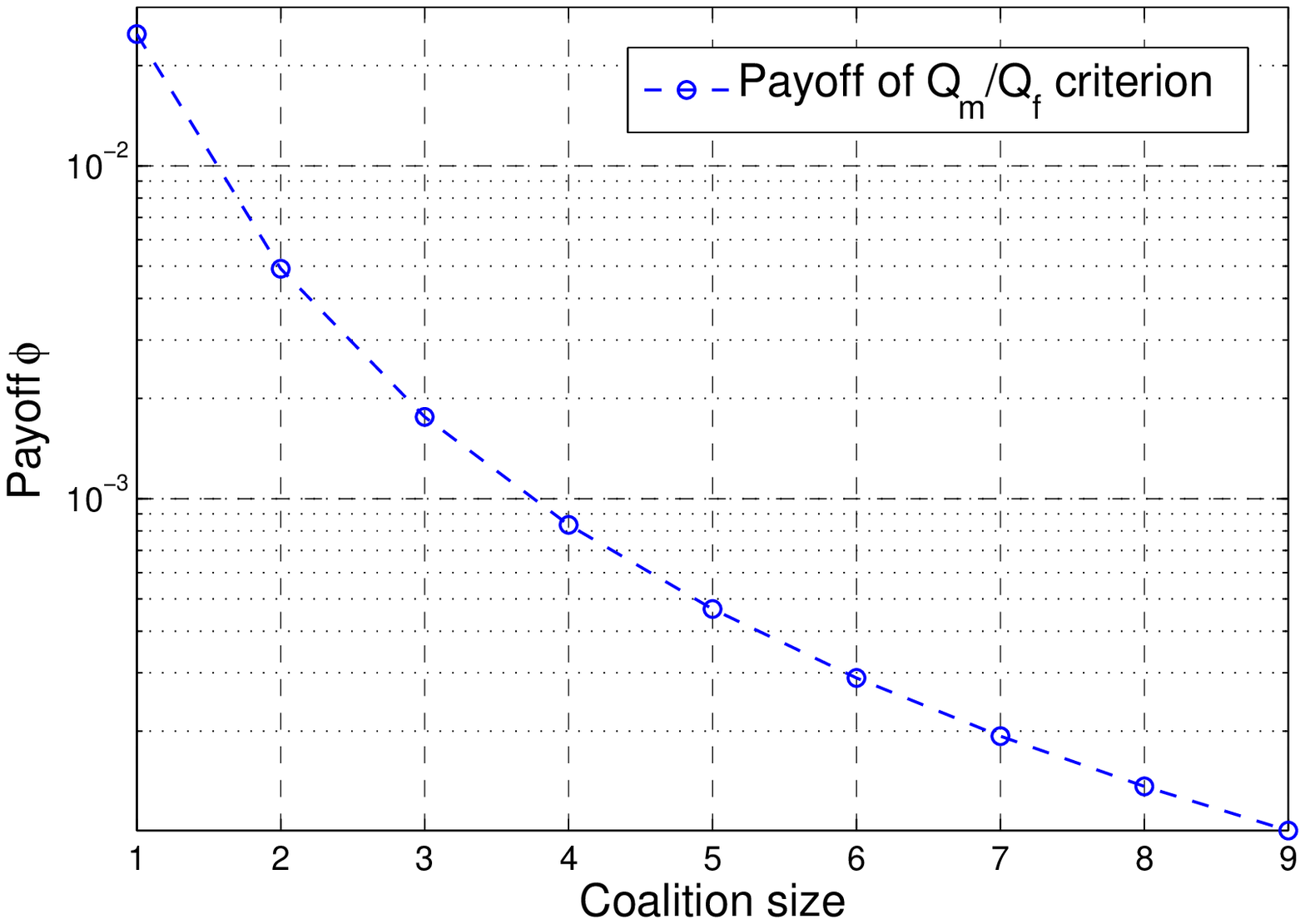}}
\caption{Coalition payoff as a function of coalition size for both the $Q_m+Q_f$ and $Q_m/Q_f$ criteria.}
\label{phi_ab} %% label for entire figure
\end{figure}

For any player $j\ne i, j \in \mathcal{R}_i$, the payoff from coalition $\mathcal{R}_i$ is defined by the marginal utility due to player $j$'s joining, given by:
\begin{equation} \label{payoffj}
\phi_j(\mathcal{R}_i) = U(\mathcal{R}_i) - U(\mathcal{R}_i \setminus \{j\}) = U(|\mathcal{R}_i|) - U(|\mathcal{R}_i|-1),
\end{equation}
the payoff of player $i$ is the remaining utility after coalition $\mathcal{R}_i$ pays all the other members, i.e.,
\begin{equation} \label{payoffi}
\phi_i(\mathcal{R}_i) = U(|\mathcal{R}_i|) - \left(|\mathcal{R}_i|-1\right)\left[U(|\mathcal{R}_i|) - U(|\mathcal{R}_i|-1)\right].
\end{equation}
Due to the monotone-increasing property and the concavity of the utility function, as given by (\ref{property1}) and (\ref{property2}), all the payoffs are positive, and only determined by the coalition size $|\mathcal{R}_i|$. The numerical results in Fig.~\ref{phi_ab} show that $\phi_j(|\mathcal{R}_i|), j \ne i$ is a decreasing convex function in both criteria, i.e.,
\begin{equation} \label{property11}
\phi_j(|\mathcal{R}_x|) < \phi_j(|\mathcal{R}_y|), \text{ with } |\mathcal{R}_x|>|\mathcal{R}_y|,
\end{equation}
and
\begin{equation} \label{property22}
\phi_j(|\mathcal{R}_x|-1) - \phi_j(|\mathcal{R}_x|) < \phi_j(|\mathcal{R}_y|-1) - \phi_j(|\mathcal{R}_y|), \text{ with } |\mathcal{R}_x|>|\mathcal{R}_y|.
\end{equation}
For any given coalitional structure $\mathcal{CS}$, the total payoff of player $i \in \mathcal{N}$ is then given by:
\begin{equation} \label{payofft}
\Phi_i(\mathcal{CS}) = \sum\limits_{i \in \mathcal{R}_j, \mathcal{R}_j \in \mathcal{CS}}\phi_i(\mathcal{R}_j).
\end{equation}
Note that $\Phi_i(\mathcal{CS})$ is only determined by the sizes of the coalitions that player $i$ participates in and the total payoff of all players is equal to the social welfare.

\begin{definition}\label{DefOCF-game}
The proposed \emph{OCF-game} is defined by the pair $(\mathcal{N},U)$, where $\mathcal{N}$ is the set of players, and $U: 2^N \rightarrow \mathbb{R}$, given by~(\ref{utility}), is the utility function. For any given coalitional structure $\mathcal{CS}$, the individual payoff of SU $i \in \mathcal{N}$ is $\Phi_i(\mathcal{CS})$, given by~(\ref{payofft}).
\end{definition}

In the proposed OCF-game model, the first step of the suboptimal solution is strictly captured by the local resource limitations, and the second steps (\ref{Opt12}) and (\ref{Opt22}) are captured by the utility function (\ref{utilityOrigin}). Therefore, the centralized optimization problems (\ref{Opt1}) and (\ref{Opt2}) can be cast as the proposed OCF-game where the players choose their strategies in a distributed manner so as to maximize their own payoffs. As the individual payoffs increase, the social welfare also increases, and, in this case, the objective functions in (\ref{Opt1}) and (\ref{Opt2}) approach closer to their optimal values.

\subsection{Algorithm based on Overlapping Coalition Formation}

We propose a DCS algorithm that consists of three stages: (1) the neighbor discovery (ND) stage, (2) the coalition formation (CF) stage, and (3) the threshold decision (TD) stage. In the ND stage, each SU discovers nearby SUs as well as the distance to each of its neighbors. In the CF stage, the SUs communicate with each other via the control channel (reporting channel) and decide which SUs to report, or equally, which coalitions to join. In the TD stage, each SU decides its local sensing threshold using a local method. After the completeness of all the three stages, the SUs can perform DCS as described in the system model, with the reporting matrix and the local sensing thresholds given by the DCS algorithm. The proposed DCS algorithm based on overlapping coalition formation is shown in Table~\ref{OCF}.

In the ND stage, a number of existing ND algorithms can be applied over the control channel~\cite{HJC-2006,SRAC-2009}. We assume the neighbors within distance $\sqrt[\mu]{ (\kappa P_{SU}) / (\gamma_0 \sigma_u^2) }$ are discovered, so that the received power at any SU is above $\kappa_0$ when its neighbor transmits at full power $P_{SU}$. The set of SU $i$'s neighbors is denoted by $\mathcal{N}_i$. Note that the concept of neighbor is reciprocal. The distance $d_{i,j}$ between any two neighboring SU $i$ and SU $j$ is known by both ends.

In the CF stage, we propose a coalition formation algorithm based on the proposed OCF-game model. First, each SU initializes its state by joining as many coalitions as possible, i.e., each SU joins coalitions from the nearest to the farthest as long as its resource is sufficient. Formally, for SU $i\in \mathcal{N}$ with neighbors $n_{1}, n_{2}, \ldots, n_{L}, L = |\mathcal{N}_i|$, we assume $d_{i,n_{j}} \le d_{i,n_{j+1}}, \forall 1\le j < L $. Then, SU $i$ sequentially joins coalitions $\mathcal{R}_{n_1}, \mathcal{R}_{n_2}, \ldots,\mathcal{R}_{n_l}$ until the remaining power or bandwidth resource is insufficient for the next coalition $\mathcal{R}_{n_{l+1}}$, or it already joins all the nearby coalitions ($l=L$). Note that SU $i$ naturally belongs to coalition $\mathcal{R}_{i}$ in all cases without contributing any power or bandwidth resource.

After the initialization, the SUs iteratively adjust their report-to sets $\mathcal{S}_i, i \in \mathcal{N}$ in a random order, so as to maximize their individual total payoff. Given the current coalitional structure $\mathcal{CS} = \{ \mathcal{R}_{1}, \mathcal{R}_{2}, \ldots, \mathcal{R}_{N}\}$, the best strategy of SU $i$ is formulated as:
\begin{subequations} \label{knapsack}
\begin{align}
\max\limits_{\mathcal{S}_i \subseteq \mathcal{N}_i} &~ \sum\limits_{j \in \mathcal{S}_i} { \phi_i(\mathcal{R}_j \cup \{i\}) }, \\
s.t. &~ \sum\limits_{j \in \mathcal{S}_i}{ \frac{\gamma_0 \sigma_{u}^2}{\kappa d_{i,j}^{-\mu}} } \le P_{SU}, \\
     &~ |\mathcal{S}_i| \theta_0 \le \theta_{SU}.
\end{align}
\end{subequations}
Problem (\ref{knapsack}) is a knapsack problem with an extra constraint on the number of objects, which in most general cases is NP-complete~\cite{HW-2006}. Here, we propose a ``switch" operation for SUs to adjust their report-to sets, after which the total payoff of the considered SU is guaranteed to increase. The main idea of switch operation is to leave one low-paying coalition and join another high-paying coalition, as long as the remaining power can cover the possible extra consumption. The convergence of switch operations is proved in the next subsection.

\begin{definition}\label{DefSwitch}
Given the coalitional structure $\mathcal{CS} = \{ \mathcal{R}_{1}, \mathcal{R}_{2}, \ldots, \mathcal{R}_{N}\}$, a \emph{switch} operation of player $i \in \mathcal{N}$ with remaining power $P_i$ is defined by a pair $(\mathcal{R}_x,\mathcal{R}_y)$ that satisfies:
\begin{equation}
\frac{\gamma_0 \sigma_{u}^2}{\kappa d_{i,y}^{-\mu}} - \frac{\gamma_0 \sigma_{u}^2}{\kappa d_{i,x}^{-\mu}} \le P_i,
\end{equation}
and
\begin{equation}
\phi_i(\mathcal{R}_y \cup \{i\}) > \phi_i(\mathcal{R}_x),
\end{equation}
where $x,y \in \mathcal{N}_i$ and $i \in \mathcal{R}_x, i \notin \mathcal{R}_y$. For any SU $i \in \mathcal{N}$, a switch operation $(\mathcal{R}_x,\mathcal{R}_y)$ implies that SU $i$ leaves coalition $\mathcal{R}_x$ and joins coalition $\mathcal{R}_y$.
\end{definition}

In the TD stage, each coalition $\mathcal{R}_i$ seeks to find the optimal threshold vector $\mathbf{\Lambda}(\mathcal{R}_i)$ in (\ref{optLambda}), so as to achieve the coalition utility as defined in (\ref{utilityOrigin}). However, an SU may belong to multiple coalitions and the optimal threshold of one coalition is not necessarily the optimal threshold of the other coalitions. Therefore, we need a threshold decision algorithm for each SU to determine its practical sensing threshold. Generally speaking, this local threshold should be a function of the optimal thresholds $\lambda(\mathcal{R}_j)$ for all $\mathcal{R}_j$ including $i$. In the $Q_m/Q_f$ criterion, in order to guarantee the false alarm probability, the SU should choose the maximum value of all the expected thresholds, i.e., $\max\nolimits_{j|i \in \mathcal{R}_j} \lambda_{b}(|\mathcal{R}_j|)$. In the $Q_m+Q_f$ criterion, there are no constraints for false alarm or missed detection probabilities. Considering that each coalition represents the sensing performance of an SU, for fairness, the SU should choose the average value of all the expected thresholds, i.e., $[\sum\nolimits_{j|i \in \mathcal{R}_j} \lambda_{a}(|\mathcal{R}_j|)] [\sum\nolimits_{j|i \in \mathcal{R}_j} 1]^{-1}$, where $[\sum\nolimits_{j|i \in \mathcal{R}_j} 1]$ is the number of coalitions that SU $i$ joins. Thus, for any final coalitional structure $\mathcal{CS}_f = \{\mathcal{R}_1, \mathcal{R}_2, \ldots, \mathcal{R}_N\}$, the local sensing threshold of SU $i \in \mathcal{N}$ is formally given by:
\begin{equation} \label{lambda}
\lambda_i =
\begin{cases}
\left[\sum\limits_{j|i \in \mathcal{R}_j} \lambda_{a}(|\mathcal{R}_j|)\right] \left[\sum\limits_{j|i \in \mathcal{R}_j} 1\right]^{-1},  &~ \text{$Q_m+Q_f$ criterion} \\
\max\limits_{j|i \in \mathcal{R}_j} \lambda_{b}(|\mathcal{R}_j|),  &~ \text{$Q_m/Q_f$ criterion}
\end{cases}
\end{equation}
where  $\lambda_{a}(\cdot)$ and $\lambda_{b}(\cdot)$ are given in Appendix A and B, respectively.

\begin{table}[!t]
\renewcommand{\arraystretch}{2.0}
\caption{DCS Algorithm based on Overlapping Coalition Formation}
\label{OCF}
\centering
\begin{tabular}{p{150mm}}

\hline

\textbf{Neighbor Discovery Stage}

Each SU $i \in \mathcal{N}$ discovers the SUs within distance $\sqrt[\mu]{ (\kappa P_{SU}) / (\gamma_0 \sigma_u^2) }$, the set of which is denoted by $\mathcal{N}_i$, and also the distance $d_{i,j}$ for any neighbor $j \in \mathcal{N}_i$. \\

\textbf{Coalition Formation Stage}

Each SU joins as many coalitions as possible by informing the corresponding SUs about its joning and the initial coalitional structure is given by $\mathcal{CS}_0$.

\begin{algorithmic} [1]
\STATE $\mathcal{CS} \leftarrow \mathcal{CS}_0$ $\%$ initial coalitional structure
\WHILE{SU $i$ has a switch operation $(\mathcal{R}_x,\mathcal{R}_y)$ as defined in Definition \ref{DefSwitch}}
\STATE SU $i$ informs SU $x$ that it leaves coalition $\mathcal{R}_x$.
\STATE SU $x$ informs SUs $j \ne x, j \in \mathcal{R}_x \backslash \{i\}$ that SU $i$ leaves coalition $\mathcal{R}_x$.
\STATE The corresponding SUs update their information about coalition $\mathcal{R}_x \leftarrow \mathcal{R}_x \backslash \{i\}$.
\STATE SU $i$ informs SU $y$ that it joins coalition $\mathcal{R}_y$.
\STATE SU $y$ informs SUs $j \ne y, j \in \mathcal{R}_y$ that SU $i$ joins coalition $\mathcal{R}_y$.
\STATE The corresponding SUs update their information about coalition $\mathcal{R}_y \leftarrow \mathcal{R}_y \cup \{i\}$.
\ENDWHILE
\STATE $\mathcal{CS}_f \leftarrow \mathcal{CS}$ $\%$ final coalitional structure
\end{algorithmic} \\

\textbf{Threshold Decision Stage}

For each SU $i \in \mathcal{N}$, the local sensing threshold $\lambda_i$ is given by (\ref{lambda}) with the current coalitional structure $\mathcal{CS}_f$.  \\

\hline

\end{tabular}
\end{table}

\subsection{Convergence and Overhead}

\begin{theorem} \label{theorem1}
In the proposed OCF-game with any initial coalitional structure $\mathcal{CS}_0$, the network converges to a final coalitional structure $\mathcal{CS}_f$ within $\lceil E/\varepsilon \rceil$ switch operations, where $E = \sum\nolimits_{i \in \mathcal{N}} {U(|\mathcal{N}_i|)} - \sum\nolimits_{\mathcal{R}_i \in \mathcal{CS}_0} {U(|\mathcal{R}_i|)}$ and $\varepsilon = 2U(N-1) - U(N) - U(N-2)$.
\end{theorem}

\begin{IEEEproof}
For any current coalitional structure $\mathcal{CS}$, the utilities of coalitions $\mathcal{R}_x$ and $\mathcal{R}_y$ are changed after switch operation $(\mathcal{R}_x,\mathcal{R}_y)$ of SU $i$, while the utilities of the other coalitions remain the same. For coalition $\mathcal{R}_x$, its size decreases from $|\mathcal{R}_x|$ to $|\mathcal{R}_x|-1$ and its utility decreases from $U(|\mathcal{R}_x|)$ to $U(|\mathcal{R}_x|-1)$. For coalition $\mathcal{R}_y$, its size increases from $|\mathcal{R}_y|$ to $|\mathcal{R}_y|+1$ and its utility increases from $U(|\mathcal{R}_y|)$ to $U(|\mathcal{R}_y|+1)$. Thus, the social welfare of the new coalitional structure $\mathcal{CS}'$ is given by:
\begin{align} \label{convergence}
\Upsilon(\mathcal{CS}') = & \Upsilon(\mathcal{CS}) - \left[ U(|\mathcal{R}_x|) + U(|\mathcal{R}_y|) \right] + \left[ U(|\mathcal{R}_x|-1) + U(|\mathcal{R}_y|+1) \right] \nonumber \\
 = & \Upsilon(\mathcal{CS}) + \left[U(|\mathcal{R}_y|+1) - U(|\mathcal{R}_y|)\right] - \left[U(|\mathcal{R}_x|) - U(|\mathcal{R}_x|-1)\right] \nonumber \\
 = & \Upsilon(\mathcal{CS}) + \phi_i(|\mathcal{R}_y \cup \{i\}|) - \phi_i(|\mathcal{R}_x|) \nonumber \\
 > & \Upsilon(\mathcal{CS})
\end{align}
Inequality (\ref{convergence}) shows that a switch operation always increases the social welfare. Since the payoff function is a convex decreasing function, as given in (\ref{property11}) and (\ref{property22}), we have $\Upsilon(\mathcal{CS}') - \Upsilon(\mathcal{CS}) = \phi_i(|\mathcal{R}_y \cup \{i\}|) - \phi_i(|\mathcal{R}_x|) \ge \phi_i(N-1) - \phi_i(N) = 2U(N-1) - U(N) - U(N-2)$. Thus, we have a lower bound of the marginal increase of social welfare due to a single switch operation $\varepsilon = 2U(N-1) - U(N) - U(N-2)$. Also, the coalition utility is an increasing function, as given in (\ref{property1}), we have an upper bound of social welfare when each coalition $\mathcal{R}_i \subseteq \mathcal{N}_i$ reaches its largest size $|\mathcal{N}_i|$, given by $\sum\nolimits_{i \in \mathcal{N}} {U(|\mathcal{N}_i|)}$. Thus, the gap of social welfare between $\mathcal{CS}_0$ and $\mathcal{CS}_f$ is limited by the upper bound $E = \sum\nolimits_{i \in \mathcal{N}} {U(|\mathcal{N}_i|)} - \sum\nolimits_{\mathcal{R}_i \in \mathcal{CS}_0} {U(|\mathcal{R}_i|)}$. Therefore, the network must converge within $\lceil E/\varepsilon \rceil$ switch operations.
\end{IEEEproof}

Traditionally, the stability of OCSs is studied by the notion of \emph{c-core}, in which an OCS is stable if no subset of players has the motivation to deviate from the current OCS and form new coalitions among themselves \cite{CEMJ-2010}. However, the notion \emph{c-core} is based on the assumption that the deviators~(players who remove their contribution from some of their coalitions) are untrustworthy and all coalitions should punish them by giving no payoff to them. In our proposed OCF-game, the players do not exhibit this property. In contrast, for our game, the deviators will not suffer any punishment. Thus, we need to define new notions to characterize the stability of the final OCS in the proposed algorithm.

\begin{definition}\label{DefTransferStable}
In the proposed OCF-game, OCS $\mathcal{CS}$ is \emph{switch-stable} if there does not exist a switch operation $(\mathcal{R}_x,\mathcal{R}_y)$ for any SU $i \in \mathcal{N}$ as defined in Definition~\ref{DefSwitch}.
\end{definition}

For the proposed algorithm given in Table~\ref{OCF}, we directly have:
\begin{lemma}
The final coalitional structure $\mathcal{CS}_f$ resulting from the algorithm in Table~\ref{OCF} is switch-stable.
\end{lemma}

In general, the final coalitional structure $\mathcal{CS}_f$ is not the optimal solution. Also, the specific form of $\mathcal{CS}_f$ greatly depends on the sequence that the SUs perform switch operations and it is generally not unique. However, we still have the following proposition.

\begin{proposition} \label{proposition2}
For any given CR network, let $\mathcal{CS}_{opt}$ denote the optimal coalitional structure with the highest social welfare $\Upsilon(\mathcal{CS}_{opt})$, and let $\mathcal{CS}_0$ and $\mathcal{CS}_f$ denote the initial and final coalitional structures in the proposed overlapping algorithm. We have:
\begin{equation}
\frac{\Upsilon(\mathcal{CS}_f)}{\Upsilon(\mathcal{CS}_{opt})}
\ge
\frac{\sum\nolimits_{\mathcal{R}_i \in \mathcal{CS}_0}{U(|\mathcal{R}_i|)}} {NU(\lceil \sum\nolimits_{\mathcal{R}_i \in \mathcal{CS}_0}{|\mathcal{R}_i}| / N \rceil)}.
\end{equation}
\end{proposition}

\begin{IEEEproof}
Since the utility function $U(\cdot)$ is an increasing concave function, as given in (\ref{property1}) and (\ref{property2}), the optimal social welfare satisfies:
\begin{equation} \label{prosposition21}
\Upsilon(\mathcal{CS}_{opt}) = \sum\limits_{\mathcal{R}_i \in \mathcal{CS}_{opt}}U(|\mathcal{R}_i|) \le NU(\lceil \sum\limits_{\mathcal{R}_i \in \mathcal{CS}_{opt}}{|\mathcal{R}_i|} / N \rceil).
\end{equation}

In the proposed overlapping algorithm, as we noted, each SU joins as many coalitions as possible in the initialization period of the CF stage. Therefore, the initial coalitional structure $\mathcal{CS}_0$ has the largest sum coalition size among all the feasible coalitional structures. Note that a switch operation does not change the sum size of the involved coalitions. We have:
\begin{equation} \label{prosposition22}
\sum\limits_{\mathcal{R}_i \in \mathcal{CS}_f}{|\mathcal{R}_i|} = \sum\limits_{\mathcal{R}_i \in \mathcal{CS}_0}{|\mathcal{R}_i|} \ge \sum\limits_{\mathcal{R}_i \in \mathcal{CS}_{opt}}{|\mathcal{R}_i|}.
\end{equation}

Since $U(\cdot)$ is an increasing function, as given in (\ref{property1}), by substituting (\ref{prosposition22}) into (\ref{prosposition21}), we have:
\begin{equation} \label{prosposition23}
\Upsilon(\mathcal{CS}_{opt}) \le NU(\lceil \sum\limits_{\mathcal{R}_i \in \mathcal{CS}_0}{|\mathcal{R}_i|} / N \rceil).
\end{equation}

Note that after a switch operation, the social welfare strictly increases. We have:
\begin{equation} \label{prosposition24}
\Upsilon(\mathcal{CS}_f) \ge \Upsilon(\mathcal{CS}_0) = \sum\limits_{\mathcal{R}_i \in \mathcal{CS}_0}{U(|\mathcal{R}_i|)}.
\end{equation}

Combining (\ref{prosposition23}) and (\ref{prosposition24}), we have:
\begin{equation}
\frac{\Upsilon(\mathcal{CS}_f)}{\Upsilon(\mathcal{CS}_{opt})}
\ge
\frac{\sum\nolimits_{\mathcal{R}_i \in \mathcal{CS}_0}{U(|\mathcal{R}_i|)}} {NU(\lceil \sum\nolimits_{\mathcal{R}_i \in \mathcal{CS}_0}{|\mathcal{R}_i|} / N \rceil)}.
\end{equation}
\end{IEEEproof}

Proposition~\ref{proposition2} shows that, in the proposed algorithm, the relative performance of the final coalitional structure $\mathcal{CS}_f$, compared with the optimal coalitional structure, is guaranteed to be above a certain threshold. This threshold only depends on the initial coalitional structure $\mathcal{CS}_0$ given by the initialization process in the coalition formation stage. For a given CR network, the initialization process generates a unique coalitional structure $\mathcal{CS}_0$. Thus, the threshold is only determined by the network parameters, and therefore, the relative performance, compared with the optimal solution, is guaranteed.

The overhead required for practically implementing the algorithm in Table \uppercase\expandafter{\romannumeral1} mainly relates to the stage in which the SUs initialize their states as well as when a switch operation is performed. We assume an SU's identity can be represented by $\tau$ bits. Note that each coalition corresponds to a particular SU. A coalition's identity also requires $\tau$ bits. For the message that SU $i$ leaves or joins coalition $\mathcal{R}_j$, by ignoring the $1$ bit to distinguish ``leave" and ``join", we can transmit this message in a packet of $2\tau$ bits.

In the initialization of coalitional structure $\mathcal{CS}_0$, each SU $i \in \mathcal{N}$ receives the information from SU $j\ne i, j\in\mathcal{R}_i$ that SU $j$ joins coalition $\mathcal{R}_i$. Thus, the overhead for initialization is given by:
\begin{equation} \label{overheadOCF1}
T_{init}(\mathcal{CS}_0) = \sum\limits_{\mathcal{R}_i \in \mathcal{CS}_0} 2(|\mathcal{R}_i|-1)\tau.
\end{equation}

For performing a switch operation $(\mathcal{R}_x,\mathcal{R}_y)$, SU $i$ informs SU $x$ that SU $i$ it wishes to leave coalition $\mathcal{R}_x$, and informs SU $y$ that it will join coalition $\mathcal{R}_y$. Then, SU $x$ and SU $y$ update their coalition information by informing their coalition members about SU $i$'s joining or leaving coalition $\mathcal{R}_x$ or $\mathcal{R}_y$. Thus, the overhead of switch operation $(\mathcal{R}_x,\mathcal{R}_y)$ is given by:
\begin{equation} \label{overheadOCF2}
T_{switch}(\mathcal{R}_x,\mathcal{R}_y) = 4\tau + 2\tau(|\mathcal{R}_x|-2) + 2\tau(|\mathcal{R}_y| - 1) = 2\tau(|\mathcal{R}_x| + |\mathcal{R}_y| - 1).
\end{equation}

The coalition size is approximately $\mathcal{O}(N)$. Thus, the overhead of the initialization period is $\mathcal{O}(N^2)$, and the overhead of a single switch operation is $\mathcal{O}(N)$. Note that the network converges within $\lceil E/\varepsilon \rceil$ switch operations, as given by Theorem~\ref{theorem1}. The worst-case overhead is approximately $\mathcal{O}(N^2) + \lceil E/\varepsilon \rceil \mathcal{O}(N)$.

%%%%%%%%%%%%%%%%%%%%%%%%%%%%%%%%%%%%%%%%%%%%%%%%%%%%%%%%%%%%%%%%%%%%%%%%%%%%%%%%%%%%%%%%%%%%%%%%%%%%%%%%%%%%%%%%
\section{DCS based on Non-overlapping Coalition Formation Games}%
%%%%%%%%%%%%%%%%%%%%%%%%%%%%%%%%%%%%%%%%%%%%%%%%%%%%%%%%%%%%%%%%%%%%%%%%%%%%%%%%%%%%%%%%%%%%%%%%%%%%%%%%%%%%%%%%

In this section, we extend the popular non-overlapping CF-game model for cooperative sensing that is proposed in~\cite{HNSBH-2011} while considering the power and bandwidth constraints and allowing the utility to reflect the $Q_m/Q_f$ criterion as well as the $Q_m+Q_f$ criterion. Here, we reconsider the CF-game model with the newly defined coalition utility, and then, we point out its limitations when compared to the more general OCF-game model of Section~\uppercase\expandafter{\romannumeral4}.

\subsection{Non-overlapping CF-game Model}

In the non-overlapping CF-game, the players are also the SUs $\mathcal{N} = \{1,2,\ldots,N\}$ with power resource $P_{SU}$ and bandwidth resource $\theta_{SU}$. The players form disjoint non-overlapping coalitions and the coalitional structure $\mathcal{CS}$ is a \emph{partition} of $\mathcal{N}$. Each player $i \in \mathcal{N}$ that belongs to coalition $\mathcal{C} \subseteq \mathcal{N}$ reports to the players in the same coalition by contributing power $\sum\nolimits_{j \ne i, j \in \mathcal{C}} (\kappa_0 \sigma_{u}^2) / (A d_{i,j}^{-\mu})$ and bandwidth $(|\mathcal{C}|-1) \theta_0$. Thus, each SU in $\mathcal{C}$ can receive the sensing data of all SUs in $\mathcal{C}$, and the utility of coalition $\mathcal{C}$ is thus:
\begin{equation} \label{utilityCF}
V(\mathcal{C}) = \sum\limits_{i \in \mathcal{C}} {U(\mathcal{C})} = |\mathcal{C}| ~ U(|\mathcal{C}|),
\end{equation}
where $U(|\mathcal{C}|)$ is given by (\ref{utility}). Unlike the OCF-game, the coalition in non-overlapping CF-game represents the sum performance of all its coalition members. To achieve the utility defined in (\ref{utilityCF}), we have the optimal threshold vector given by:
\begin{equation} \label{optLambdaCF}
\mathbf{\Lambda}(\mathcal{C}) =
\begin{cases}
(\lambda_{a}(|\mathcal{C}|), \lambda_{a}(|\mathcal{C}|), \ldots, \lambda_{a}(|\mathcal{C}|))_{1\times |\mathcal{C}|}, &~ \text{$Q_m+Q_f$ criterion} \\
(\lambda_{b}(|\mathcal{C}|), \lambda_{b}(|\mathcal{C}|), \ldots, \lambda_{b}(|\mathcal{C}|))_{1\times |\mathcal{C}|}, &~ \text{$Q_m/Q_f$ criterion}
\end{cases}
\end{equation}
where $\lambda_{a}(\cdot)$ and $\lambda_{b}(\cdot)$ are given in Appendix A and B, respectively.

The social welfare is also defined as the sum utility of all the coalitions, given by
\begin{equation}
\Xi(\mathcal{CS}) = \sum\limits_{\mathcal{C} \in \mathcal{CS}} |\mathcal{C}| U(|\mathcal{C}|) = \sum\limits_{j \in \mathcal{N} | j \in \mathcal{C}} U(|\mathcal{C}|).
\end{equation}
Since $U(|\mathcal{C}|)$ reflects the sensing performance of each SU in $\mathcal{C}$, then, the defined social welfare also reflects the network sensing performance. As similar as the OCF-game model, due to the monotone-increasing property of $U(\cdot)$, the network performs better as the average coalition size increases. Also, due to the increasing power cost for joining a larger coalition, the grand coalition may not always form.

We assume that the utility of each coalition is equally distributed to each coalition member, and the individual payoff of any player $i \in \mathcal{N}$ is then given by:
\begin{equation} \label{payoffCF}
\Psi_i(\mathcal{CS}) = \psi_i(\mathcal{C}) = U(|\mathcal{C}|),
\end{equation}
where $i \in \mathcal{C}$ and $\mathcal{C} \in \mathcal{CS}$. Note that the coalitions are completely disjoint and each SU belongs to only one coalition. The total payoff of an SU is the payoff from the coalition it belongs to. Naturally, the total payoff of all SUs is equal to the defined social welfare.

\begin{definition}\label{DefCF-game}
The proposed \emph{CF-game} is defined by the pair $(\mathcal{N},V)$, where $\mathcal{N}$ is the set of players, and $V: 2^N \rightarrow \mathbb{R}$, given by (\ref{utilityCF}), is the utility function. For any given coalitional structure $\mathcal{CS}$, the individual payoff of SU $i \in \mathcal{N}$ is $\Psi_i(\mathcal{CS})$, given by~(\ref{payoffCF}).
\end{definition}

Compared with the OCF-game model defined in Definition \ref{DefOCF-game}, the non-overlapping CF-game model also captures the suboptimal solution by the local resource limitations and its newly defined utility function (\ref{utilityCF}). Moreover, the optimal sensing threshold given by (\ref{optLambdaCF}) is more practical since the coalitions are disjoint and each SU belongs to only one coalition. Therefore, the increase of individual payoff, or equally the increase of social welfare, means an equal increase of the objective functions in (\ref{Opt1}) and (\ref{Opt2}).

However, the non-overlapping CF-game model imposes extra limitations on the reporting structure due to the non-overlapping assumption. From the perspective of an OCF-game, any coalition $\mathcal{C}$ in the non-overlapping CF-game model represents $|\mathcal{C}|$ identical coalitions in the OCF-game model $\mathcal{R}_i, i \in \mathcal{C}$. Thus, the non-overlapping CF-game model can be seen as a special case of the OCF-game model, in which the $N$ overlapping coalitions are classified into groups, and in each group, the coalitions are identical to a coalition consisting of the SUs that these coalitions correspond to. Next, we show the limitation of the CF-game model via a special case. In Fig.~\ref{system_model}, there are three nearby SUs $\{1,2,3\}$ and we assume each SU can only report to one SU due to the power and bandwidth constraints. In the OCF-game model, we can expect the coalitional structure $\mathcal{R}_1 = \{1,3\}, \mathcal{R}_2 = \{2,1\}, \mathcal{R}_3 = \{3,2\}$ to form. Thus, the sensing performance of the SUs are respectively given by $U(2),U(2)$, and $U(2)$. In the non-overlapping CF-game model, the network forms a structure with a two-SU coalition and a singleton, and, thus, the sensing performance will be given by $U(1),U(2)$ and $U(2)$. Clearly, the result of the OCF-game model strictly outperforms the non-overlapping CF-game model.

\subsection{Algorithm based on Non-overlapping Coalition Formation}

In CF-games, the merge-and-split algorithm is often used to achieve a stable coalitional structure \cite{SHDHB-2009,HNSBH-2011}. In this algorithm, multiple coalitions merge into one larger coalition and a single coalition split into multiple smaller coalitions, as long as the payoffs of all the involved players are increased. In the proposed CF-game, each player's payoff increases with the coalition size, as seen in (\ref{payoffCF}). Thus, the players always prefer larger coalitions and the merge-and-split algorithm degrades to the merge algorithm where the coalitions keep merging until the bandwidth or power resource is completely used for some players. The proposed DCS algorithm based on non-overlapping coalition formation is formally given in Table~\ref{CF}.

In the ND stage, we use the same method as in the overlapping case where the SUs within distance $\sqrt[\mu]{ (\kappa P_{SU}) / (\gamma_0 \sigma_u^2) }$ are discovered as the neighbors, the neighbor set of SU $i \in \mathcal{N}$ is also denoted by $\mathcal{N}_i$.

In the CF stage, we define the merge operation as follows:

\begin{definition}\label{DefMerge}
Given the coalitional structure $\mathcal{CS}$, a \emph{merge} operation in $\mathcal{CS}$ is defined by a pair $(\mathcal{C}_1,\mathcal{C}_2)$ of two disjoint coalitions that satisfies:
\begin{equation}
\sum\limits_{j \ne i, j\in \mathcal{C}_1 \cup \mathcal{C}_2} {\frac{\gamma_0 \sigma_{u}^2}{\kappa d_{i,j}^{-\mu}}} \le P_{SU}, i \in \mathcal{C}_1 \cup \mathcal{C}_2,
\end{equation}
\begin{equation}
(|\mathcal{C}_1 \cup \mathcal{C}_2|-1) \theta_0 \le \theta_{SU},
\end{equation}
where any two SUs in $\mathcal{C}_1$ and $\mathcal{C}_2$ are neighbors, i.e., $i \in \mathcal{N}_j, \forall i,j \in \mathcal{C}_1 \cup \mathcal{C}_2$.
\end{definition}

Suppose each coalition $\mathcal{C} \subset \mathcal{N}$ has a coalition head that has the complete information of all the coalition members, i.e., $\mathcal{N}_i, i \in \mathcal{C}$ and $d_{i,j}, i \in \mathcal{C}, j \in \mathcal{N}_i$. Therefore, the merge operation between two coalitions is actually performed by the two coalitions heads that represent them. Note that any feasible merge operation require all the involved players to be neighbors. Any coalition member can be chosen as the coalition head without missing any feasible merge operations. For the coalition formed by a merge operation, the coalition head is randomly chosen from the two original coalitions heads. For practical reasons, each coalition head maintains a tag parameter for each of its neighboring coalition heads. Formally, for any two neighboring heads $i$ and $j$ of coalitions $\mathcal{C}_1$ and $\mathcal{C}_2$, tags $t_{i,j} = 0$ and $t_{j,i} = 0$ represents that a merge operation $(\mathcal{C}_1,\mathcal{C}_2)$ is not feasible. A coalition head only tries the merge operations with nonzero tags. If an SU is no longer a coalition head, the corresponding tags are deleted.

In the TD stage, the optimal threshold vector $\mathbf{\Lambda}(\mathcal{C})$ for any coalition $\mathcal{C} \in \mathcal{CS}_f$ is given by (\ref{optLambdaCF}), where $\mathcal{CS}_f$ is the final coalition structure given by the CF stage. Thus, the local sensing threshold of SU $i \in \mathcal{C}, \mathcal{C} \in \mathcal{CS}_f$ is formally given by:
\begin{equation} \label{lambdaCF}
\lambda_i =
\begin{cases}
\lambda_{a}(|\mathcal{C}|),  &~ \text{$Q_m+Q_f$ criterion} \\
\lambda_{b}(|\mathcal{C}|).  &~ \text{$Q_m/Q_f$ criterion}
\end{cases}
\end{equation}
where $\lambda_{a}(\cdot)$ and $\lambda_{b}(\cdot)$ are given in Appendix A and B, respectively.

\begin{table}[!t]
\renewcommand{\arraystretch}{2.0}
\caption{DCS Algorithm based on Non-overlapping Coalition Formation}
\label{CF}
\centering
\begin{tabular}{p{150mm}}

\hline

\textbf{Neighbor Discovery Stage}

Each SU $i \in \mathcal{N}$ discovers the SUs within distance $\sqrt[\mu]{ (\kappa P_{SU}) / (\gamma_0 \sigma_u^2) }$, the set of which is denoted by $\mathcal{N}_i$, and also the distance $d_{i,j}$ for any neighbor $j \in \mathcal{N}_i$. \\

\textbf{Coalition Formation Stage}

\begin{algorithmic} [1]
\STATE $\mathcal{CS} \leftarrow \{\{1\},\{2\},\ldots,\{N\}\}$ $\%$ each SU forms a singleton
\STATE $t_{i,j} \leftarrow 1, i \in \mathcal{N}, j \in \mathcal{N}_i$ $\%$ each SU maintains a tag corresponding to a neighbor coalition head
\WHILE{SU $i \in \mathcal{C}$ has a tag $t_{i,j} = 1$}
\STATE SU $i$ sends SU $j$ the complete information of coalition $\mathcal{C}$.
\STATE SU $j \in \mathcal{C}'$ computes if $(\mathcal{C},\mathcal{C}')$ is a merge operation as defined in Definition \ref{DefMerge}.
\IF{$(\mathcal{C},\mathcal{C}')$ is a merge operation}
\STATE $\%$ $\mathcal{C}$ and $\mathcal{C}'$ merge into $\mathcal{C} \cup \mathcal{C}'$ with coalition head $j$.
\STATE SU $j$ informs SUs $k \in \mathcal{C}$ that SUs in $\mathcal{C}'$ join their coalition $\mathcal{C}$, and sets tags $t_{j,k} \leftarrow 0$.
\STATE SUs $k \in \mathcal{C}$ update their coalition information and set all their tags to zero.
\STATE SU $j$ informs SUs $k \ne j, k \in \mathcal{C}'$ that SUs in $\mathcal{C}$ join their coalition $\mathcal{C}'$, and sets tags $t_{j,k} \leftarrow 0$.
\STATE SUs $k \ne j, k \in \mathcal{C}'$ update their coalition information and set all their tags to zero.
\ELSE
\STATE SU $j$ informs SU $i$ that the trying $(\mathcal{C},\mathcal{C}')$ fails, and sets tag $t_{j,i} \leftarrow 0$.
\STATE SU $i$ sets its tag $t_{i,j} \leftarrow 0$.
\ENDIF
\ENDWHILE
\STATE $\mathcal{CS}_f \leftarrow \mathcal{CS}$ $\%$ final coalitional structure
\end{algorithmic} \\

\textbf{Threshold Decision Stage}

For any SU $i \in \mathcal{N}$, the local sensing threshold $\lambda_i$ is determined by (\ref{lambdaCF}) with the current coalitional structure $\mathcal{CS}_f$.  \\

\hline

\end{tabular}
\end{table}

\subsection{Convergence and Overhead}

The convergence of the non-overlapping coalition formation algorithm is a direct result of the defined merge rule, and follows directly from known results such as \cite{SHDHB-2009,HNSBH-2011,SHBDH-2011}. Actually, we can expect the algorithm to converge within $N$ merge operations, since each merge operation will decrease the number of coalitions by $1$. In the proposed algorithm in Table~\ref{CF}, the main source of overhead pertains to the case when a coalition tries to merge with another coalition and transmits the complete coalition information. If the merge operation is feasible and actually executed, the new coalition head updates coalition information with additional overhead, and the ``retired" coalition head informs its original members about its ``retirement" with $1$ bit information. If the merge operation is not feasible, only the $1$ bit fail information is transmitted. Here, we also assume an SU's identity requires $\tau$ bits and ignore the $1$ bit information. Note that in the CF-game model, each SU belongs to one and only one coalition. The information that SU $i$ joins coalition $\mathcal{C}$ received by SU $j \in \mathcal{C}$ can be represented by the identity of SU $i$ without causing any ambiguity.

The complete information of coalition $\mathcal{C}$ includes the information of each coalition member $i \in \mathcal{C}$, which consists of SU $i$ itself, all its neighbors in $\mathcal{N}_i$ and the corresponding distances $d_{i,j}, j\in \mathcal{N}_i$. We simply assume a distance requires $\tau$ bits. Thus, the complete information of coalition $\mathcal{C}$ is given by:
\begin{equation} \label{try}
T_{try}(\mathcal{C}) = \sum\limits_{i \in \mathcal{C}} {\left(2|\mathcal{N}_i|+1\right)}.
\end{equation}

In merge operation $(\mathcal{C},\mathcal{C}')$, coalition head $j \in \mathcal{C}'$ becomes the head of the merged coalition $\mathcal{C} \cup \mathcal{C}'$, and then, it informs the SUs in $\mathcal{C}$ about the joining of the SUs in $\mathcal{C}'$, as well as the SUs in $\mathcal{C}'$ (except for itself) about the joining of the SUs in $\mathcal{C}$. Thus, the overhead of merge operation $(\mathcal{C},\mathcal{C}')$ is given by:
\begin{equation}
T_{merge}(\mathcal{C},\mathcal{C}') = |\mathcal{C}|\times|\mathcal{C}'| + (|\mathcal{C}'|-1)\times|\mathcal{C}| = (2|\mathcal{C}'|-1) |\mathcal{C}| \tau.
\end{equation}
where SU $i$ and SU $j$ are coalition heads of coalitions $\mathcal{C}$ and $\mathcal{C}'$, respectively.

$T_{try}(\mathcal{C})$ is $\mathcal{O}(N^2)$, and $T_{merge}(\mathcal{C},\mathcal{C}')$ is $\mathcal{O}(N^2)$. In Table~\ref{CF}, in each attempt for a merge operation, at least one tag parameter is set to zero. Thus, the network converges within $N^2$ attempts, and the total overhead of trying is less than $\mathcal{O}(N^4)$. Also, the network converges within $N$ merge operations. Thus, the overhead of merge operations is $\mathcal{O}(N^3)$. Therefore, the total overhead is less than $\mathcal{O}(N^4)$.

%%%%%%%%%%%%%%%%%%%%%%%%%%%%
\section{Practical Issues}%
%%%%%%%%%%%%%%%%%%%%%%%%%%%%

In the considered DCS problem, in order to simplify our discussion and calculation, some of the practical issues are not considered. In this section, we discuss how these practical issues may effect our proposal and how we can extend our model to involve these factors.

The utility function is the fundamental characterization of an OCF-game. In the proposed OCF-game, the utility function $U(\mathcal{R}_i)$ precisely represents the sensing performance of SU $i$ in both the $Q_m+Q_f$ and $Q_m/Q_f$ criteria, as seen in (\ref{utility}). Due to the increasing monotony and concavity of $U(\mathcal{R}_i)$, the social welfare is guaranteed to be increased by each switch operation, and thus, the system sensing performance is always increasing until the network converges to a switch-stable outcome. In fact, as long as $U(\mathcal{R}_i)$ is defined as the best sensing performance we can achieve from this coalition, we can easily understand that any extra data only increases the coalition utility and the marginal improvement only decreases with the coalition size, i.e., $U(\mathcal{R}_i)$ is monotone increasing and concave.

In the system model, we assume that the PU is far away from the SUs, so that the received SNRs are the same for all SUs. However, we can replace the common received SNR $\gamma$ with $\gamma_i$ for each SU $i$, so as to consider small-scale scenarios in which the SUs have different distances to the PU. As we noted, $U(\mathcal{R}_i)$ is still monotone increasing and concave, and thus, our proposed algorithm can still work effectively, only that the calculation becomes more complex. For similar reasons, we can also extend our model to involve more practical concerns, such as more sophisticated fusion rules as in consensus-based algorithms~\cite{ALB-2011}, the spatial correlation between SUs~\cite{MS-2008, CAP-2012}, the mobility of PUs~\cite{CAP-2013}, the non-idealness of the report channel~\cite{CCPS-2013}, and even the location and time diversities of SUs in spectrum-heterogeneous systems~\cite{L-2010, ZHLYP-2011, WGWY-2013, DWWSC-2013}. For each practical issue, the utility function $U(\mathcal{R}_i)$ should be redefined to reflect such concern, but still should present the best sensing performance we can achieve for SU $i$. Note that these practical issues may complicate the utility function, and we may need to simplify $U(\mathcal{R}_i)$ to reduce the computational complexity. However, no matter how we define $U(\mathcal{R}_i)$, the properties of increasing monotony and concavity should always be guaranteed.

%%%%%%%%%%%%%%%%%%%%%%%%%%%%%%%%%%%%%%%%%%
\section{Simulation Results and Analysis}%
%%%%%%%%%%%%%%%%%%%%%%%%%%%%%%%%%%%%%%%%%%

For our simulations, we consider a network in which the SUs are randomly distributed within a $10$km$\times$$10$km square area and the PU is $D=150$km away from the square center. The path loss parameters are $\kappa=1$ and $\mu=3$, and the noise power is $\sigma_u^2 = -90$dBm. The PU transmit power $\sigma_s^2$ is set in such a way that the average received SNR at the SUs is $\gamma = -15$dB, and the number of samples at each SU is set to $N_s = 10000$. For the power and bandwidth constraints, the minimum received SNR and minimum time-frequency resource for transmitting $1$ bit are set to $\gamma_0=0$dB and $\theta_0 = 1$, respectively. In the $Q_m/Q_f$ criterion, the maximum false alarm constraint is set to $\alpha=0.1$, as recommended by the IEEE 802.22 standard~\cite{802}. The remaining parameters are varied within given ranges so as to evaluate the performance of different algorithms under different conditions. All statistical results are averaged over the random locations of the SUs via a large number of independent runs.

\subsection{Comparison of DCS algorithms}

\begin{figure}
\centering
\subfigure[Non-overlapping algorithm]{
\label{example_non} %% label for first subfigure
\includegraphics[width=2.8in]{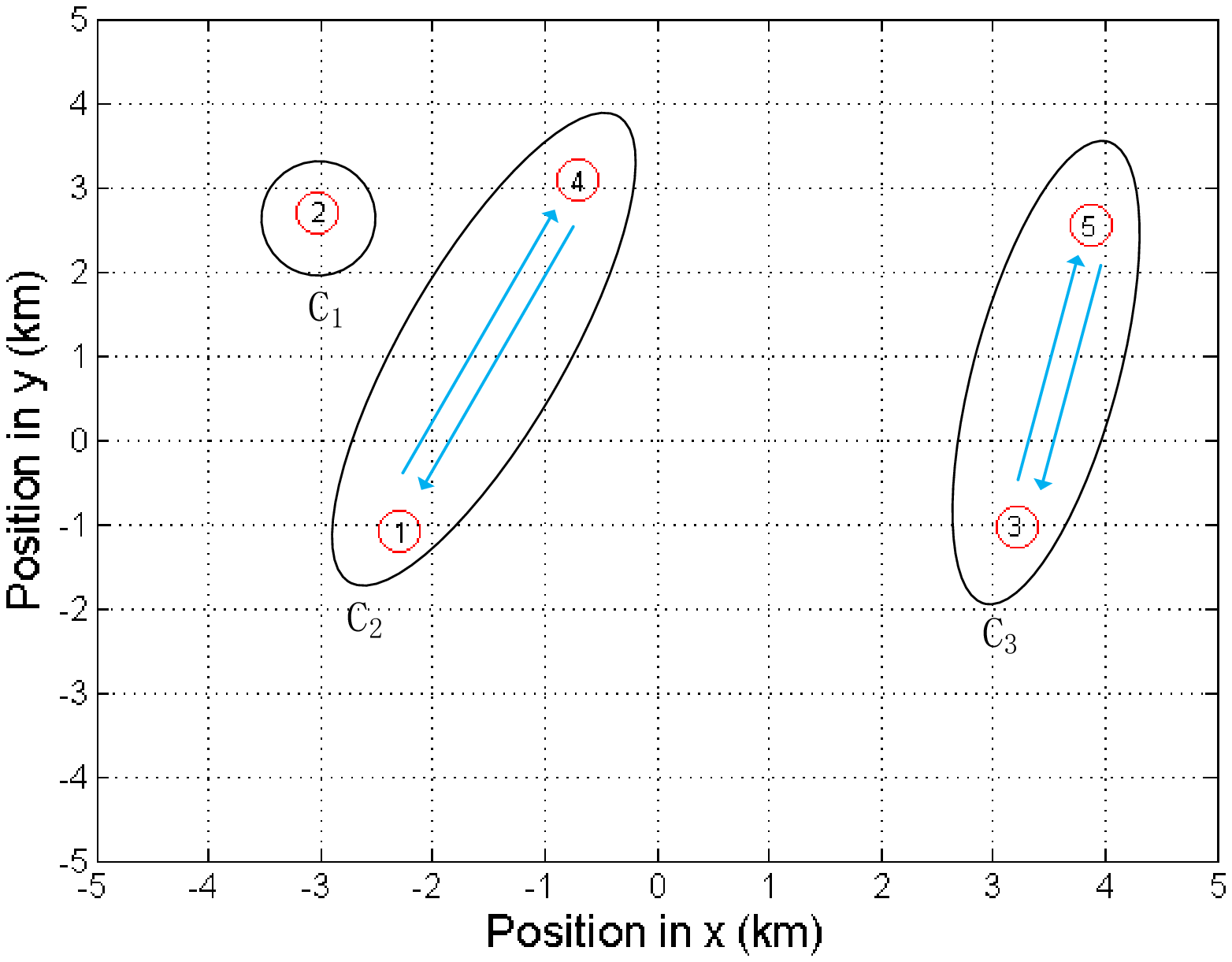}}
\hspace{-0.2in}
\subfigure[Overlapping algorithm]{
\label{example_over} %% label for second subfigure
\includegraphics[width=2.8in]{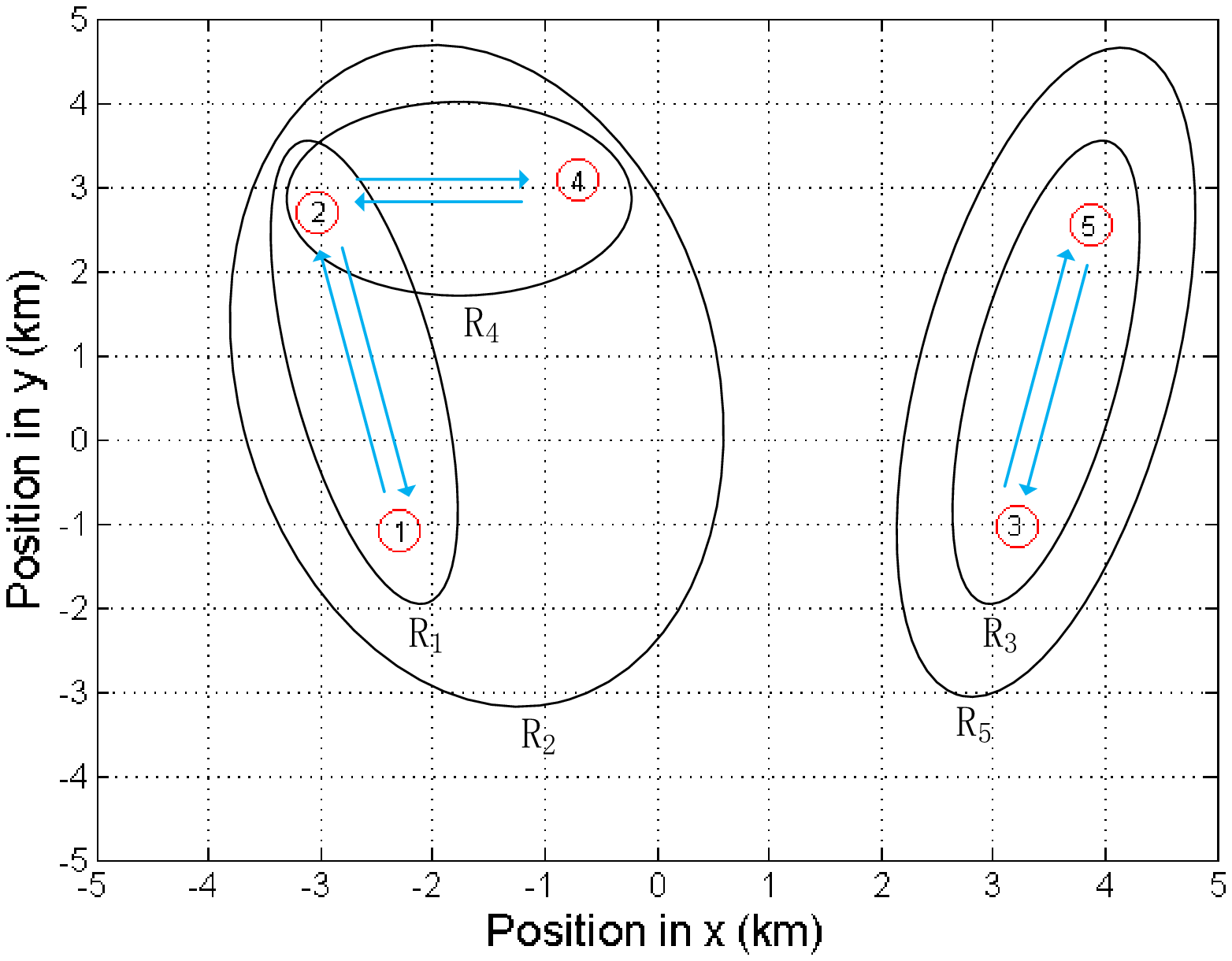}}
\caption{Snapshots of the coalitional structures resulting from both the non-overlapping and overlapping algorithms in a $5$-SU network with power constraint $P_{SU}=100$mW and bandwidth constraint $\theta_{SU}= 10$.}
\label{example} %% label for entire figure
\end{figure}

In Fig.~\ref{example}, we show the snapshots of coalitional structures of both the non-overlapping and overlapping algorithms in a $5$-SU network. The power and bandwidth constraints are set to $P_{SU} = 100$mW and $\theta_{SU} = 10$. As we see, in the non-overlapping algorithm, the SUs form a $3$-coalition non-overlapping structure $\mathcal{C}_1 = \{2\}, \mathcal{C}_2 = \{1,4\}, \mathcal{C}_3 = \{3,5\}$, and each SU reports to the SUs in the same coalition. While, in the overlapping algorithm, the SUs form a $5$-coalition overlapping structure $\mathcal{R}_1 = \{1,2\}, \mathcal{R}_2 = \{1,2,4\}, \mathcal{R}_3 = \{3,5\}, \mathcal{R}_4 = \{2,4\}, \mathcal{R}_5 = \{3,5\}$, and in each coalition, all the members report to the particular SU that the coalition corresponds to.

\begin{figure}
\centering
\subfigure[the $Q_m+Q_f$ criterion]{
\label{PerformanceN_a} %% label for first subfigure
\includegraphics[width=2.8in]{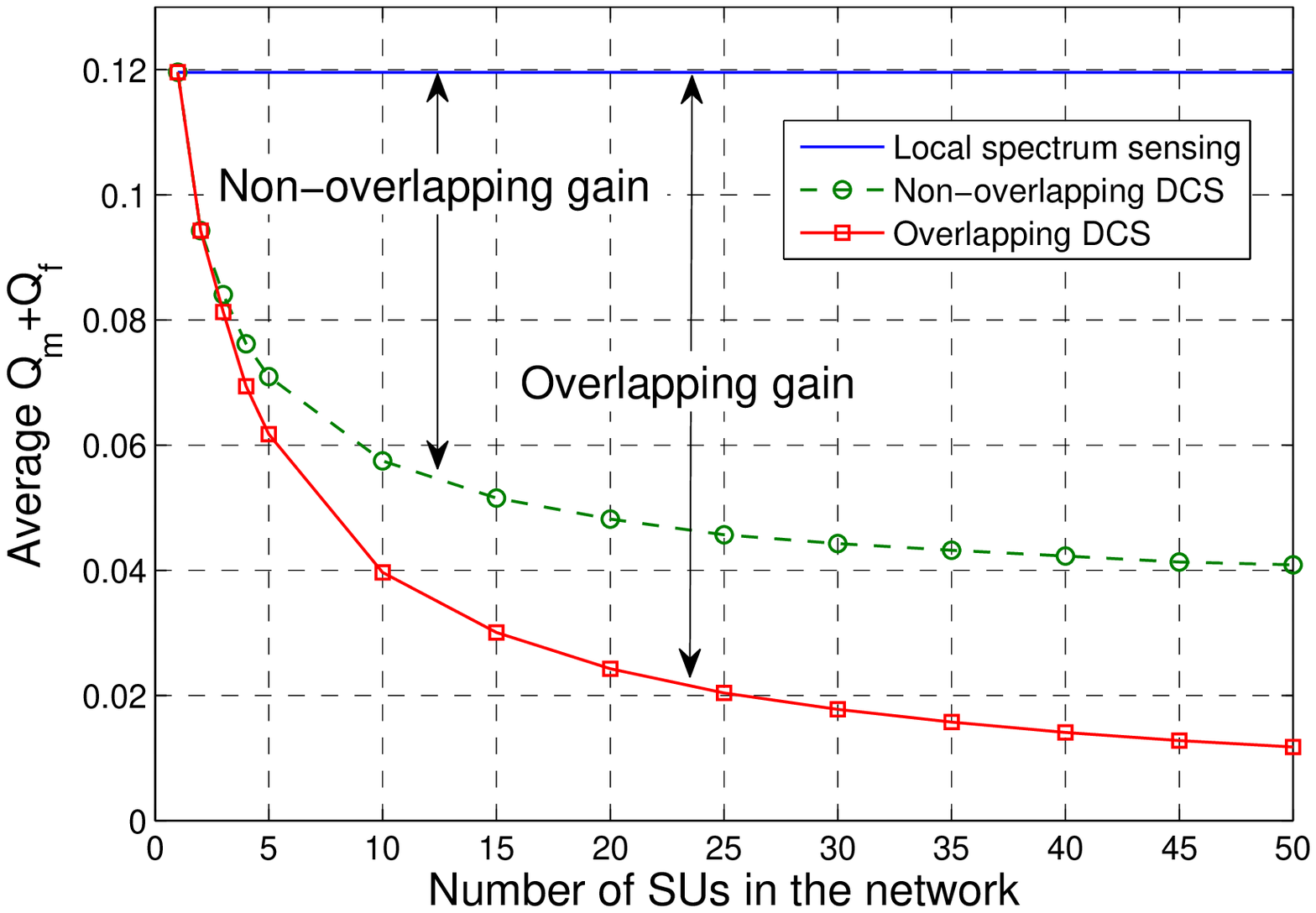}}
\hspace{-0.3in}
\subfigure[the $Q_m/Q_f$ criterion]{
\label{PerformanceN_b} %% label for second subfigure
\includegraphics[width=2.8in]{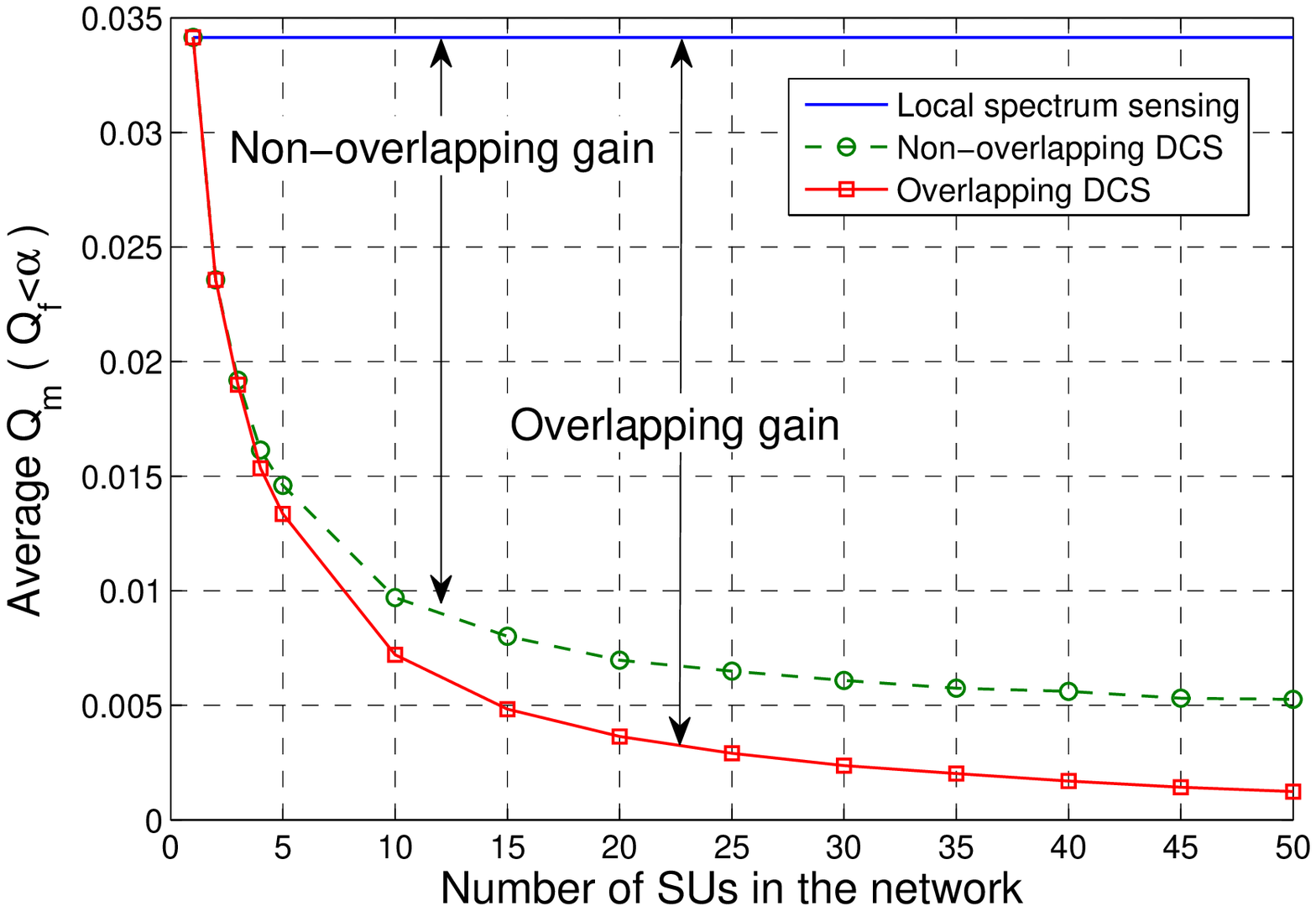}}
\caption{Network sensing performance as a function of the number of SUs $N$ with power constraint $P_{SU}=100$mW and bandwidth constraint $\theta_{SU}= 10$ in both criteria. The false alarm constraint in the $Q_m/Q_f$ criterion is $\alpha = 0.1$.}
\label{PerformanceN} %% label for entire figure
\end{figure}

In Fig.~\ref{PerformanceN}, we show the network sensing performance as a function of the network size $N$ in both the $Q_m + Q_f$ and $Q_m/Q_f$ criteria. The power and bandwidth constraints are set as $P_{SU} = 100$mW and $\theta_{SU} = 10$. It shows that, for both criteria, the proposed cooperative algorithms outperform the local spectrum sensing, and their cooperative gains increase with the network size. Also, the overlapping DCS outperforms the non-overlapping DCS in all cases, and the gap between them increases with the network size. When the network is sparse, both cooperative algorithms have similar performance. While, when the network is dense ($N=50$), the total error probability ($Q_m+Q_f$ criterion) is reduced from $0.04$ to $0.01$, which is $25\%$, and the missed detection probability ($Q_m/Q_f$ criterion) is reduced from $0.005$ to $0.001$, which is $20\%$. As the network becomes denser, each SU can cooperate with more neighbors with the same power and bandwidth resources, and thus, the average coalition size increases. In both Sections~\uppercase\expandafter{\romannumeral4} and~\uppercase\expandafter{\romannumeral5}, the network sensing performance is represented by the social welfare, which increases with the average coalition size. Therefore, the increasing network size can improve the network sensing performance by increasing the average coalition size, as seen in Fig.~\ref{PerformanceN}.

\begin{figure}[!t]
\centering
\includegraphics[width=2.8in]{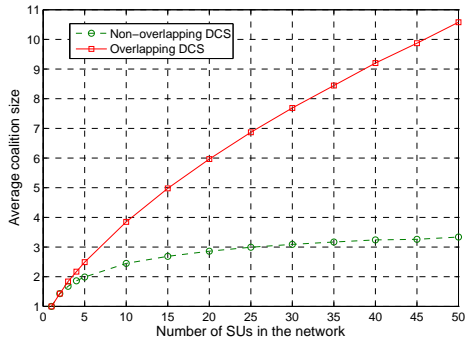}
\caption{Average coalition size as a function of the number of SUs $N$ with power constraint $P_{SU} = 100$mW and bandwidth constraint $\theta_{SU} = 10$ in either criteria.} \label{size_N}
\end{figure}

In Fig.~\ref{size_N}, we show the average coalition size as a function of network size $N$ resulting from both proposed algorithms. It clearly shows that the overlapping DCS achieves a much larger coalition size than the non-overlapping DCS, which explains the performance gap as seen in Fig.~\ref{PerformanceN}. In general, the overlapping structure provides the SUs with more flexibility on the distribution of their local resources, which encourages them to cooperate with more neighbors, and thus, increases the average coalitions size and improves the network sensing performance. Fig.~\ref{size_N} shows that the average coalition size for the overlapping case reaches a maximum of $11$ for a network with $N = 50$ SUs, while that for the non-overlapping case does not exceed $4$.

\begin{figure}
\centering
\subfigure[Non-overlapping algorithm]{
\label{PDF_a} %% label for first subfigure
\includegraphics[width=2.8in]{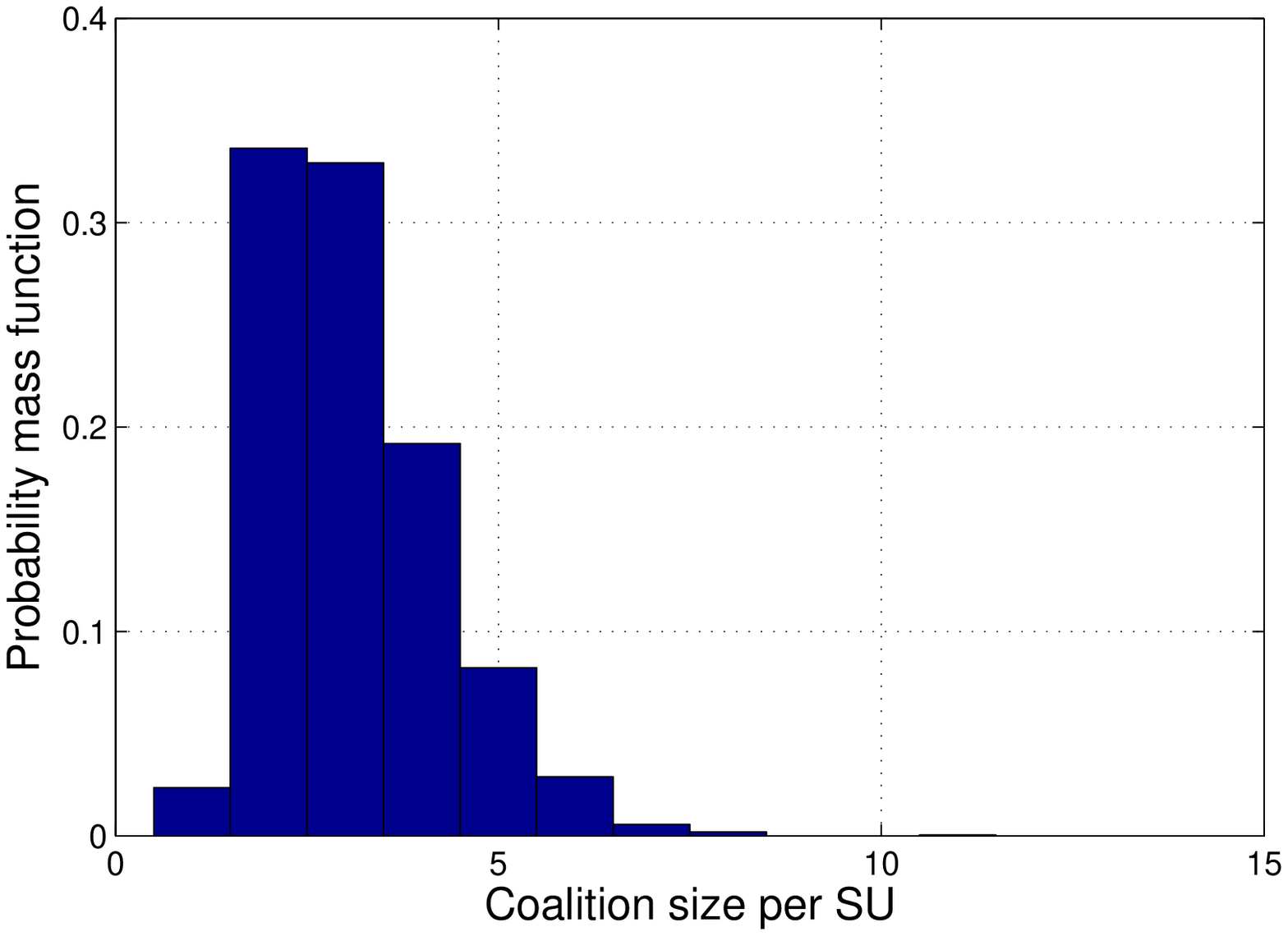}}
\hspace{-0.3in}
\subfigure[Overlapping algorithm]{
\label{PDF_b} %% label for second subfigure
\includegraphics[width=2.8in]{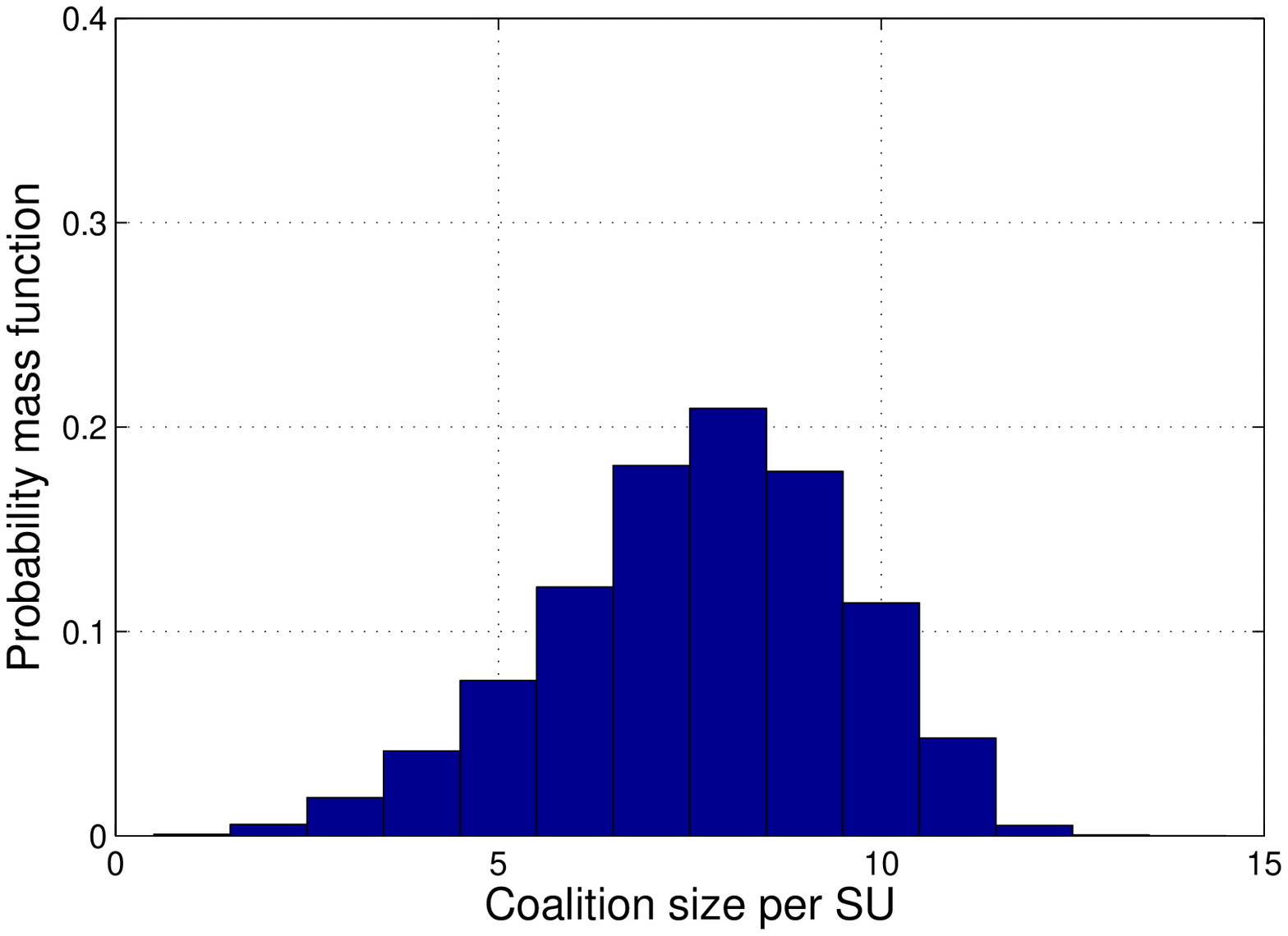}}
\caption{Probability density function of coalition size per SU for networks with $N=30$ SUs, power constraint $P_{SU}=100$mW and bandwidth constraint $\theta_{SU}= 10$ in either criteria.}
\label{PDF_ab} %% label for entire figure
\end{figure}

In Fig.~\ref{PDF_ab}, we show the probability density functions of coalition size per SU for both proposed algorithms. It shows that in the non-overlapping algorithm, the coalitions with sizes $2,3,4$ occupy about $80\%$ of all coalitions, while in the overlapping algorithm, coalitions with sizes $6,7,8,9,10$ occupy the same percentage. This is in line with the result in Fig.~\ref{size_N} which shows that the overlapping algorithm forms larger coalitions. In addition, this also implies that the variance of the sizes of the coalitions resulting from the overlapping algorithm exceeds those resulting from the non-overlapping algorithm. Thus, the SUs in the overlapping algorithm have a wider range of sensing performance.

\subsection{Power and Bandwidth Constraints}

\begin{figure}
\centering
\subfigure[the $Q_m+Q_f$ criterion]{
\label{PerformanceB_a} %% label for first subfigure
\includegraphics[width=2.8in]{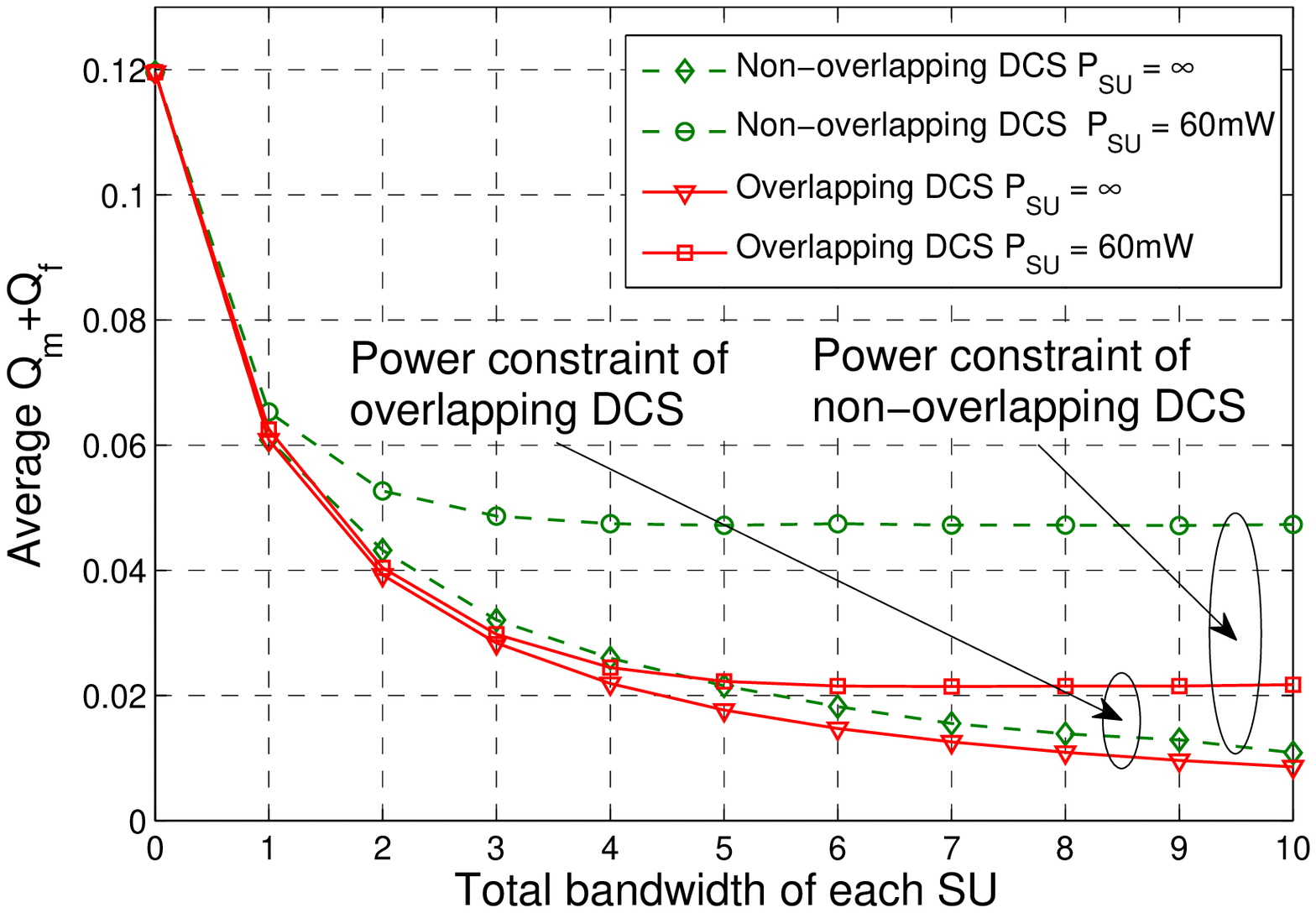}}
\hspace{-0.3in}
\subfigure[the $Q_m/Q_f$ criterion]{
\label{PerformanceB_b} %% label for second subfigure
\includegraphics[width=2.8in]{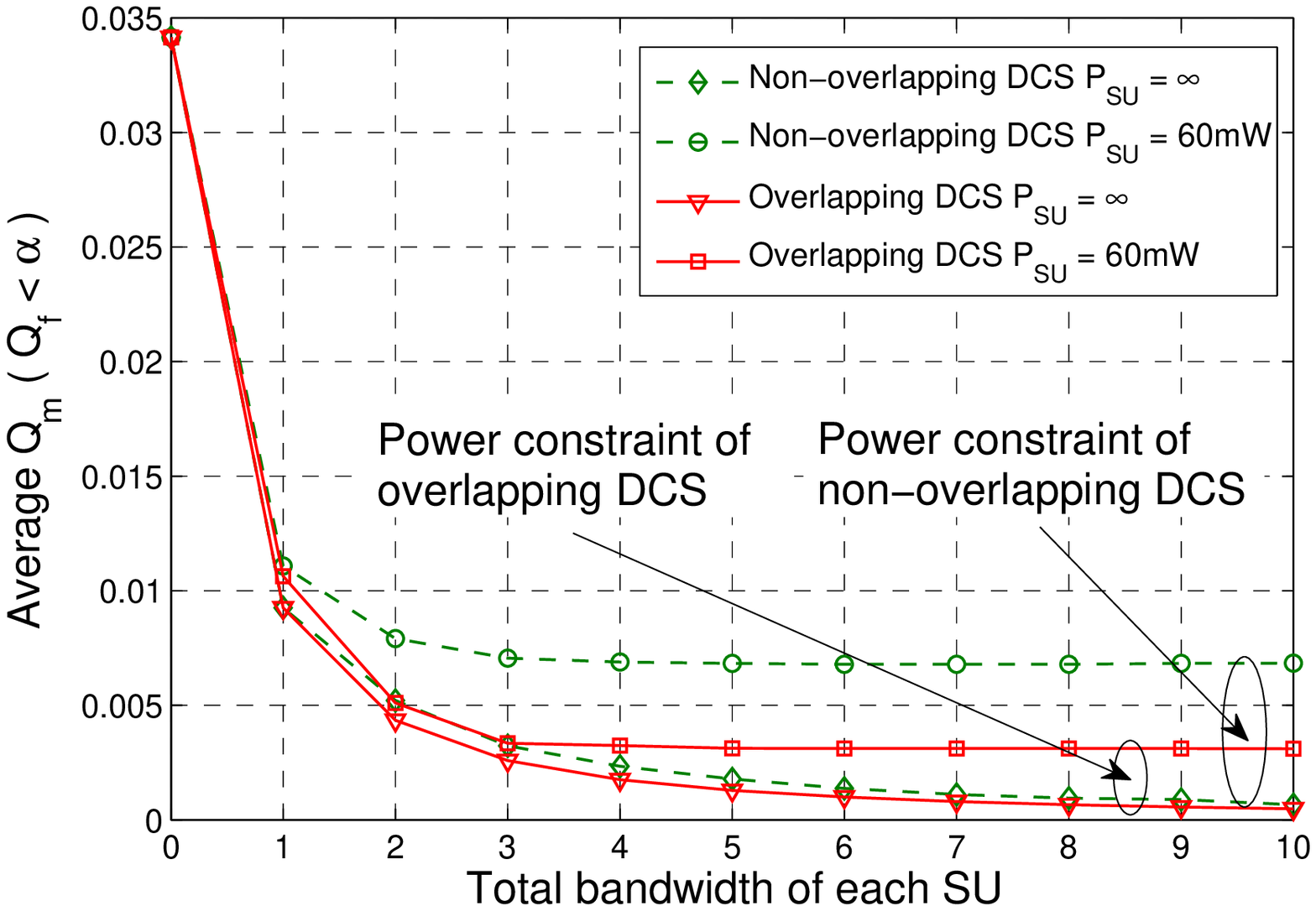}}
\caption{Network sensing performance as a function of the total bandwidth of each SU $\theta_{SU}$ with network size $N=30$. The false alarm constraint in the $Q_m/Q_f$ criterion is $\alpha = 0.1$.}
\label{PerformanceB} %% label for entire figure
\end{figure}

\begin{figure}
\centering
\subfigure[the $Q_m+Q_f$ criterion]{
\label{PerformanceP_a} %% label for first subfigure
\includegraphics[width=2.8in]{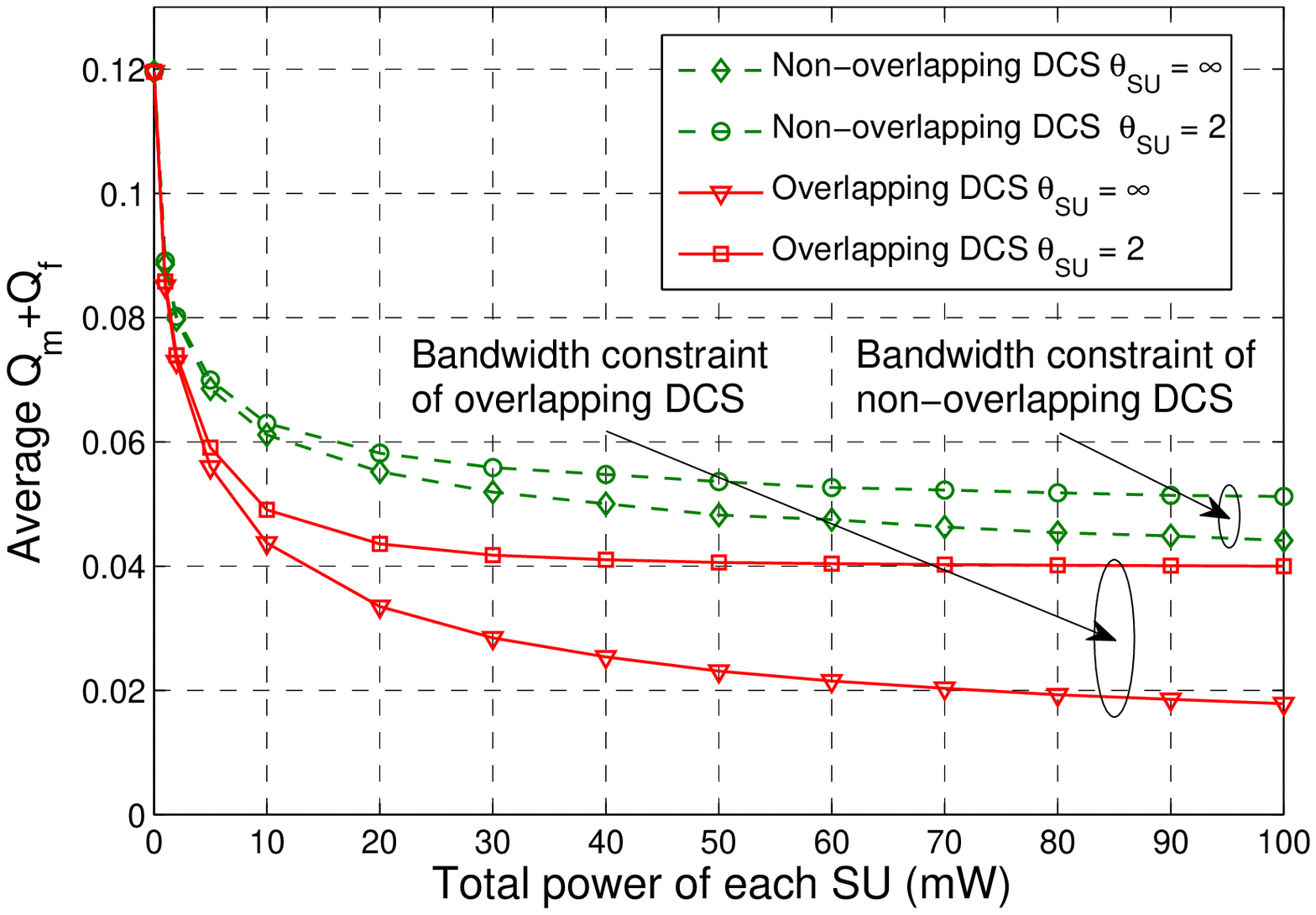}}
\hspace{-0.3in}
\subfigure[the $Q_m/Q_f$ criterion]{
\label{PerformanceP_b} %% label for second subfigure
\includegraphics[width=2.8in]{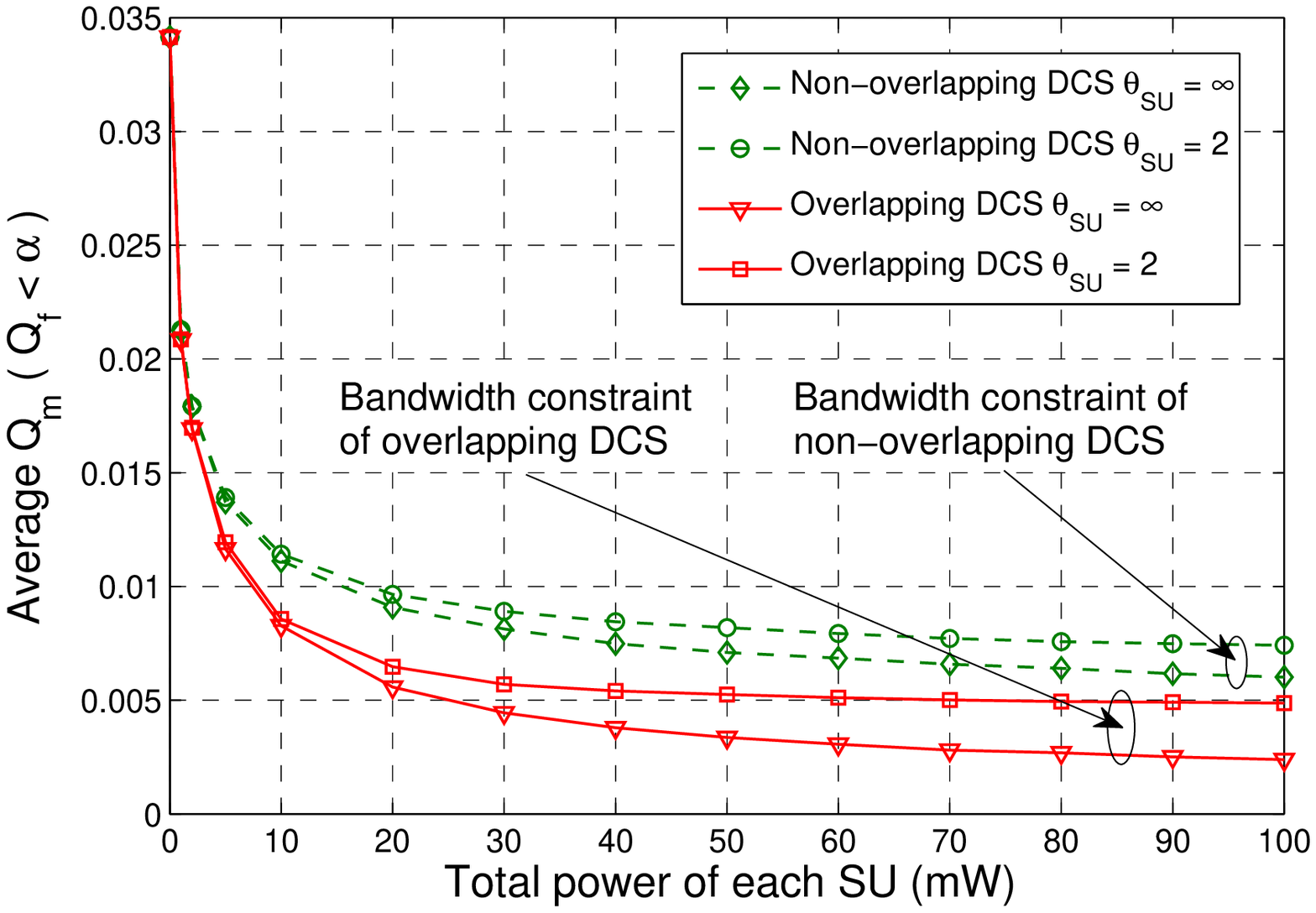}}
\caption{Network sensing performance as a function of the total power of each SU $P_{SU}$ with network size $N=30$. The false alarm constraint in the $Q_m/Q_f$ criterion is $\alpha = 0.1$.}
\label{PerformanceP} %% label for entire figure
\end{figure}

Figs.~\ref{PerformanceB} and~\ref{PerformanceP} show the network sensing performance as a function of the bandwidth $\theta_{SU}$ and the power $P_{SU}$, respectively. In Fig.~\ref{PerformanceB}, we show the curves of infinite power ($P_{SU}=\infty$) and limited power ($P_{SU}=60$mW) for each algorithm in each criterion. The gap between the two curves represents the corresponding performance decrease due to the power constraint. As similar to Fig.~\ref{PerformanceB}, we show performance decrease due to the bandwidth constraint in Fig.~\ref{PerformanceP}, by showing the curves of infinite bandwidth ($\theta_{SU}=\infty$) and limited bandwidth ($\theta_{SU}=2$). Fig.~\ref{PerformanceB} (Fig.~\ref{PerformanceP}) clearly shows that the curves of infinite power (bandwidth) decrease as the bandwidth (power) resource increases, while the curves of limited power (bandwidth) flatten out once the bandwidth (power) resource exceeds a certain threshold. Thus, the gap due to the power (bandwidth) constraint increases with the bandwidth (power) resource. In the $Q_m+Q_f$ criterion, when the bandwidth (power) is sufficient $\theta_{SU}=10$ ($P_{SU}=100$mW), the power (bandwidth) constraint increases the total error probability from $0.01$ ($0.044$) to $0.05$ ($0.052$) of the non-overlapping algorithm, and from $0.008$ ($0.02$) to $0.02$ ($0.04$) of the overlapping algorithm. In the $Q_m/Q_f$ criterion, when the bandwidth (power) is sufficient $\theta_{SU}=10$ ($P_{SU}=100$mW), the power (bandwidth) constraint increases the missed detection probability from $0.001$ ($0.006$) to $0.007$ ($0.008$) of the non-overlapping algorithm, and from $0.001$ ($0.002$) to $0.003$ ($0.005$) of the overlapping algorithm.

The behavior of the curves shown in Figs.~\ref{PerformanceB} and~\ref{PerformanceP} can be explained as follows. As previously noted, the network sensing performance is mainly determined by the average coalition size, which in general is limited by both power and bandwidth constraints. If the power (bandwidth) is infinite, the bandwidth (power) becomes the only limitation and the performance monotonously improves with the increasing bandwidth (power), as seen in the curves of infinite power (bandwidth) in Fig.~\ref{PerformanceB} (Fig.~\ref{PerformanceP}). If the power (bandwidth) is limited, it will become the major limitation when the bandwidth (power) is sufficiently large, and even if we keep increasing the bandwidth (power) resource, the performance keeps stationary, as we see in the curves of limited power (bandwidth) in Fig.~\ref{PerformanceB} (Fig.~\ref{PerformanceP}). Therefore, the gap due to the power (bandwidth) constraint increases with the bandwidth (power) resource, as seen in Fig.~\ref{PerformanceB} (Fig.~\ref{PerformanceP}).

\begin{figure}
\centering
\subfigure[Power utilization]{
\label{BN} %% label for first subfigure
\includegraphics[width=2.8in]{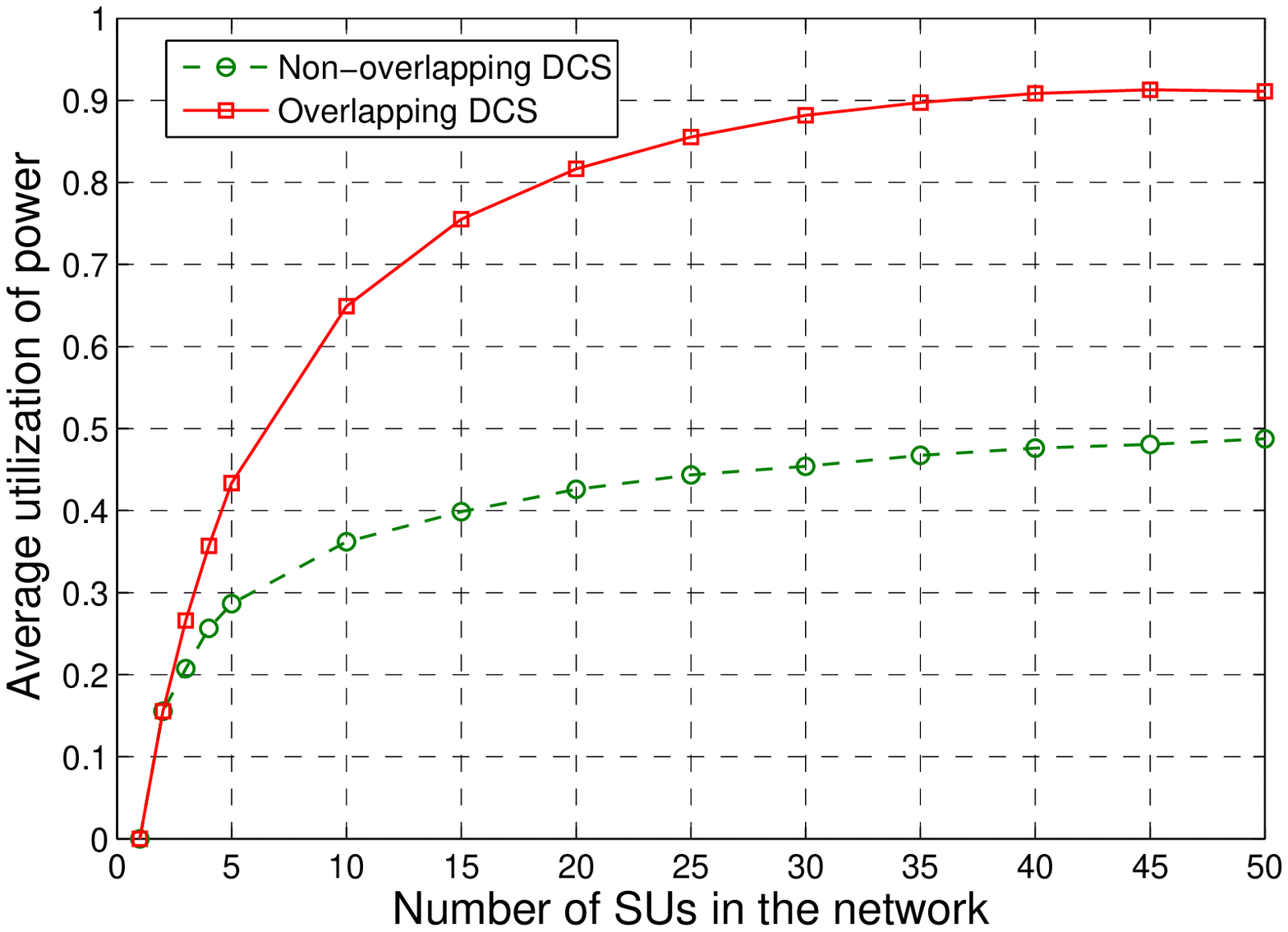}}
\hspace{-0.3in}
\subfigure[Bandwidth utilization]{
\label{PN} %% label for second subfigure
\includegraphics[width=2.8in]{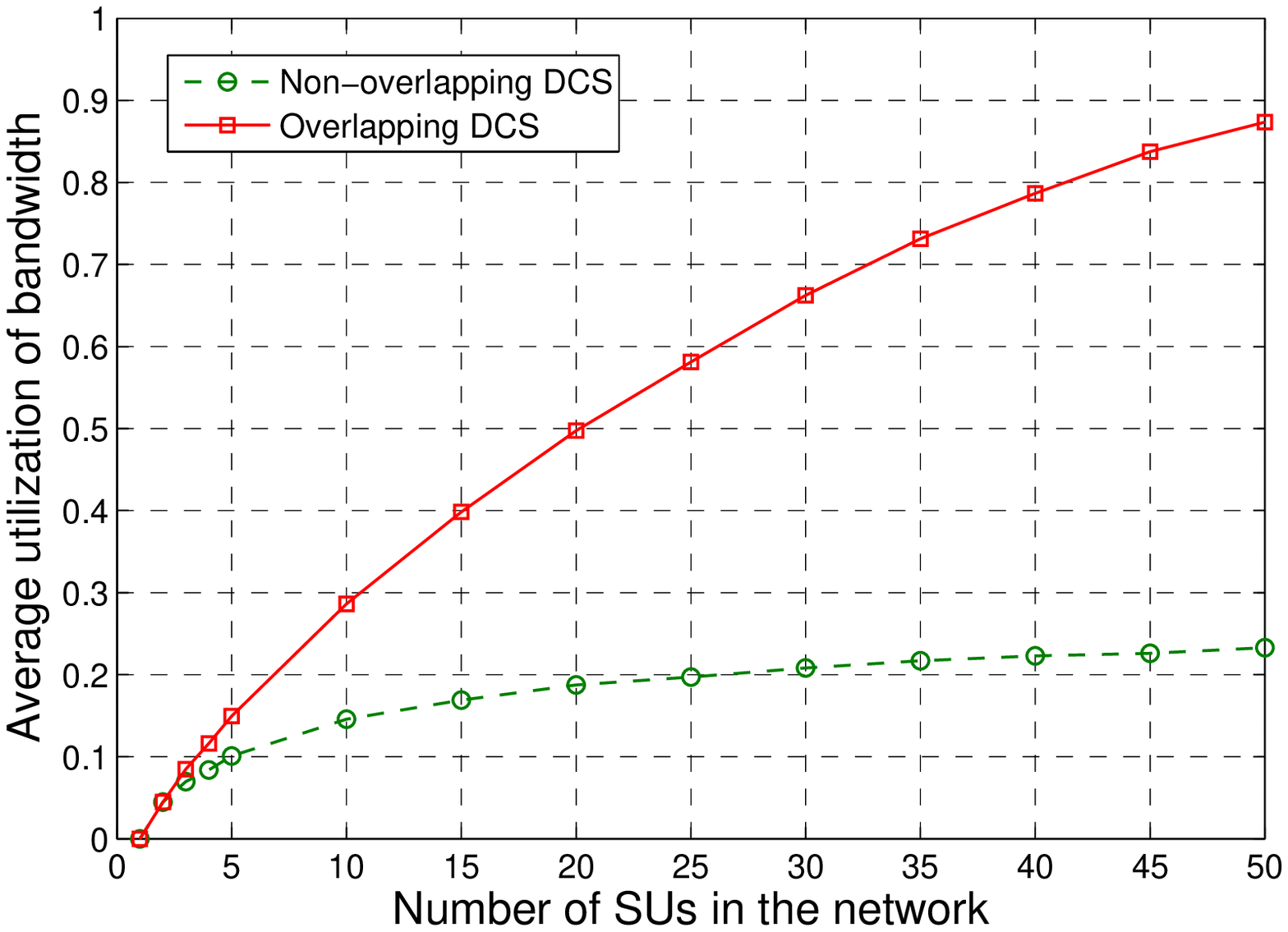}}
\caption{Average resource utilization as a function of the number of SUs $N$ with power constraint $P_{SU}=100$mW and bandwidth constraint $\theta_{SU}= 10$ in either criteria. The false alarm constraint in the $Q_m/Q_f$ criterion is $\alpha = 0.1$.}
\label{PBN} %% label for entire figure
\end{figure}

In Fig.~\ref{PBN}, we show the average resource utilization as a function of the number of SUs $N$ for both bandwidth and power resources. When the network is dense ($N=50$), $90\%$ power and $90\%$ bandwidth are utilized by the overlapping algorithm, while only $50\%$ power and $25\%$ bandwidth are utilized by the non-overlapping algorithm. As we noted, the overlapping structure provides the SUs with more flexibility on the distribution of local resources, which enables them to contribute more resources, and thus, increases the power and bandwidth utilizations, as seen in Fig.~\ref{PBN}. The higher resource utilization of the overlapping algorithm increases the average coalition size, and thus, improves the network sensing performance, as seen in Figs.~\ref{PerformanceN} and~\ref{size_N}.

\subsection{Convergence, Overhead, and Complexity}

\begin{figure}
\centering
\subfigure[]{
\label{PerformanceO_a} %% label for first subfigure
\includegraphics[width=2.8in]{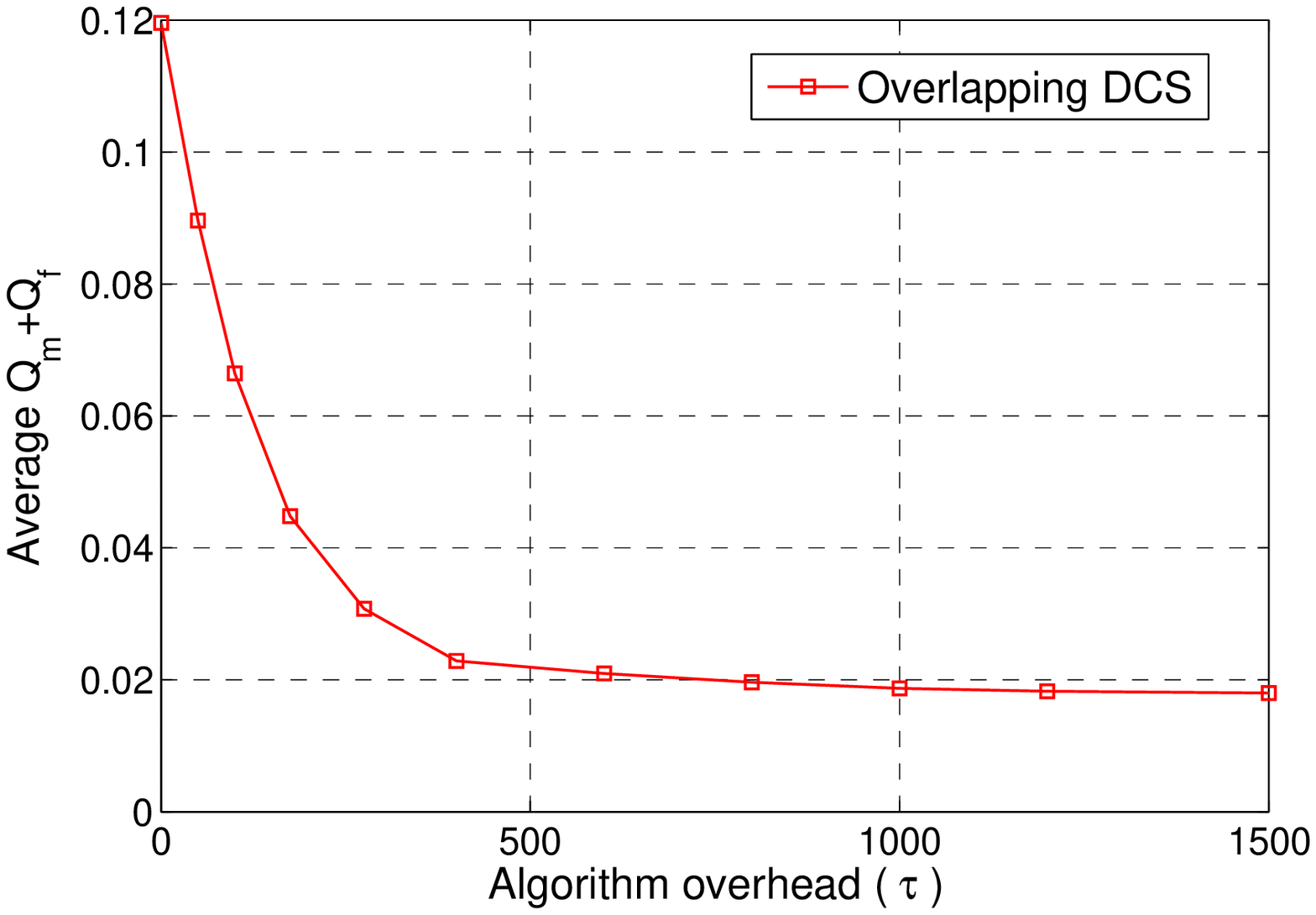}}
\hspace{-0.3in}
\subfigure[]{
\label{PerformanceO_b} %% label for second subfigure
\includegraphics[width=2.8in]{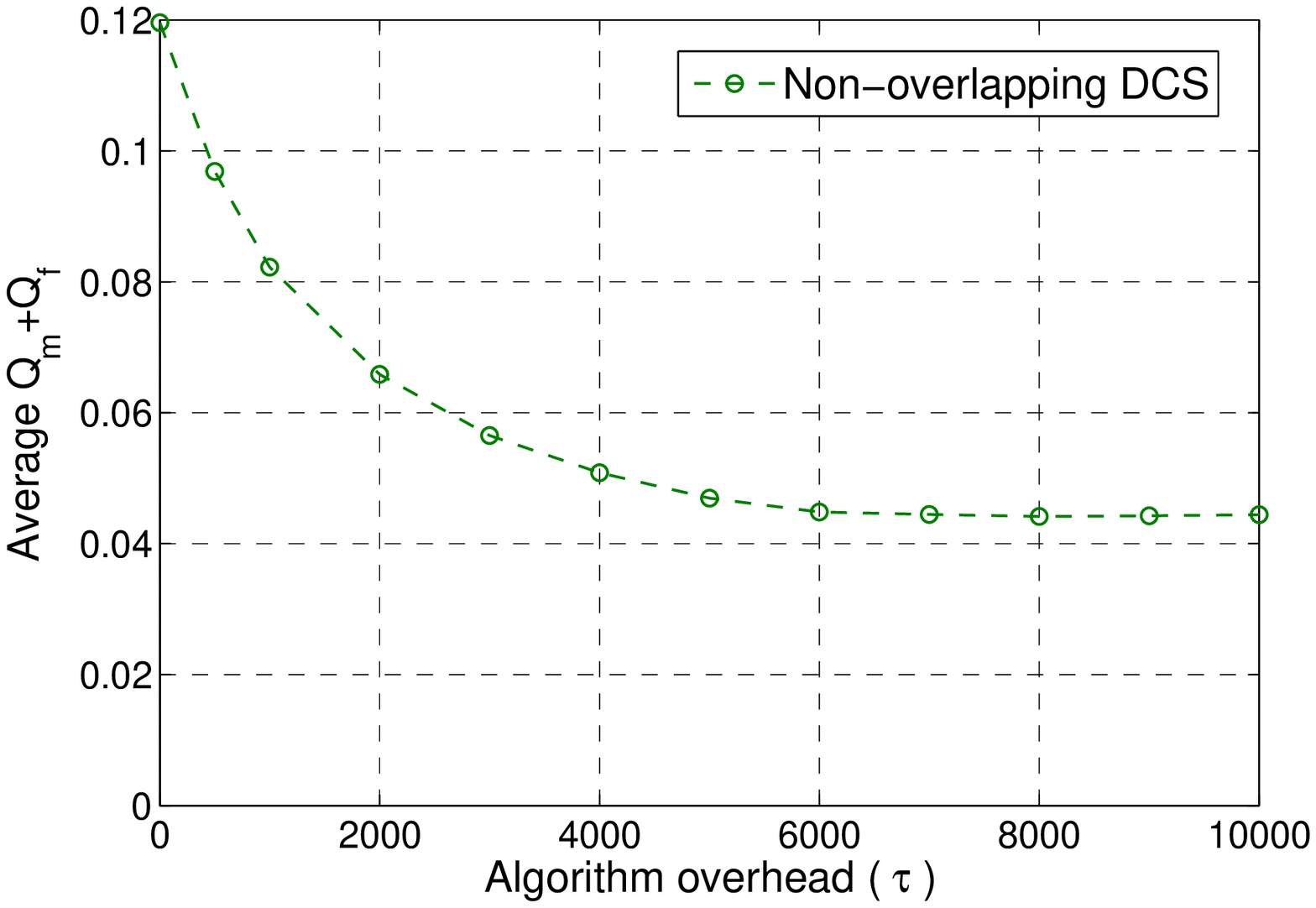}}
\subfigure[]{
\label{PerformanceO_c} %% label for second subfigure
\includegraphics[width=2.8in]{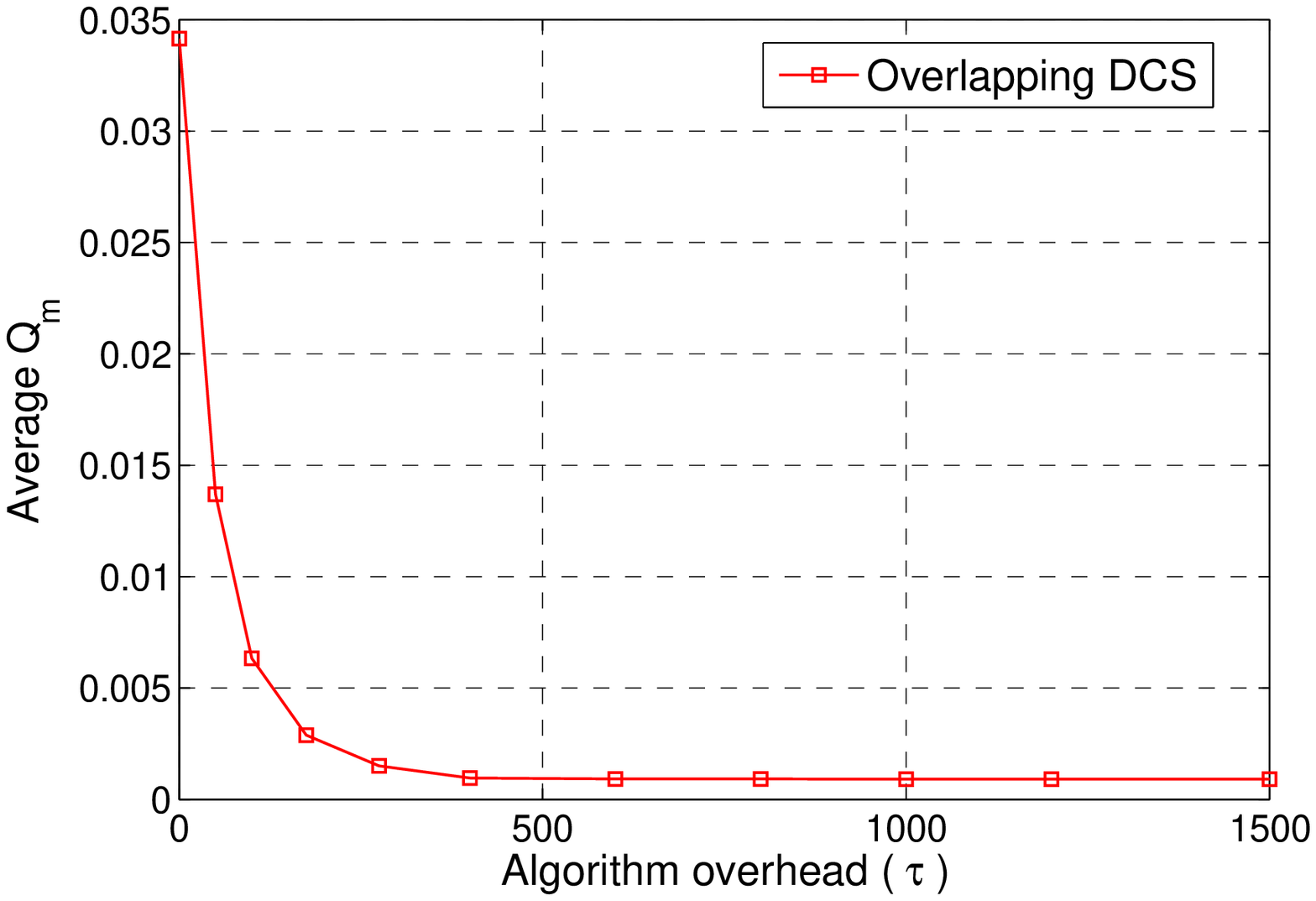}}
\hspace{-0.3in}
\subfigure[]{
\label{PerformanceO_d} %% label for second subfigure
\includegraphics[width=2.8in]{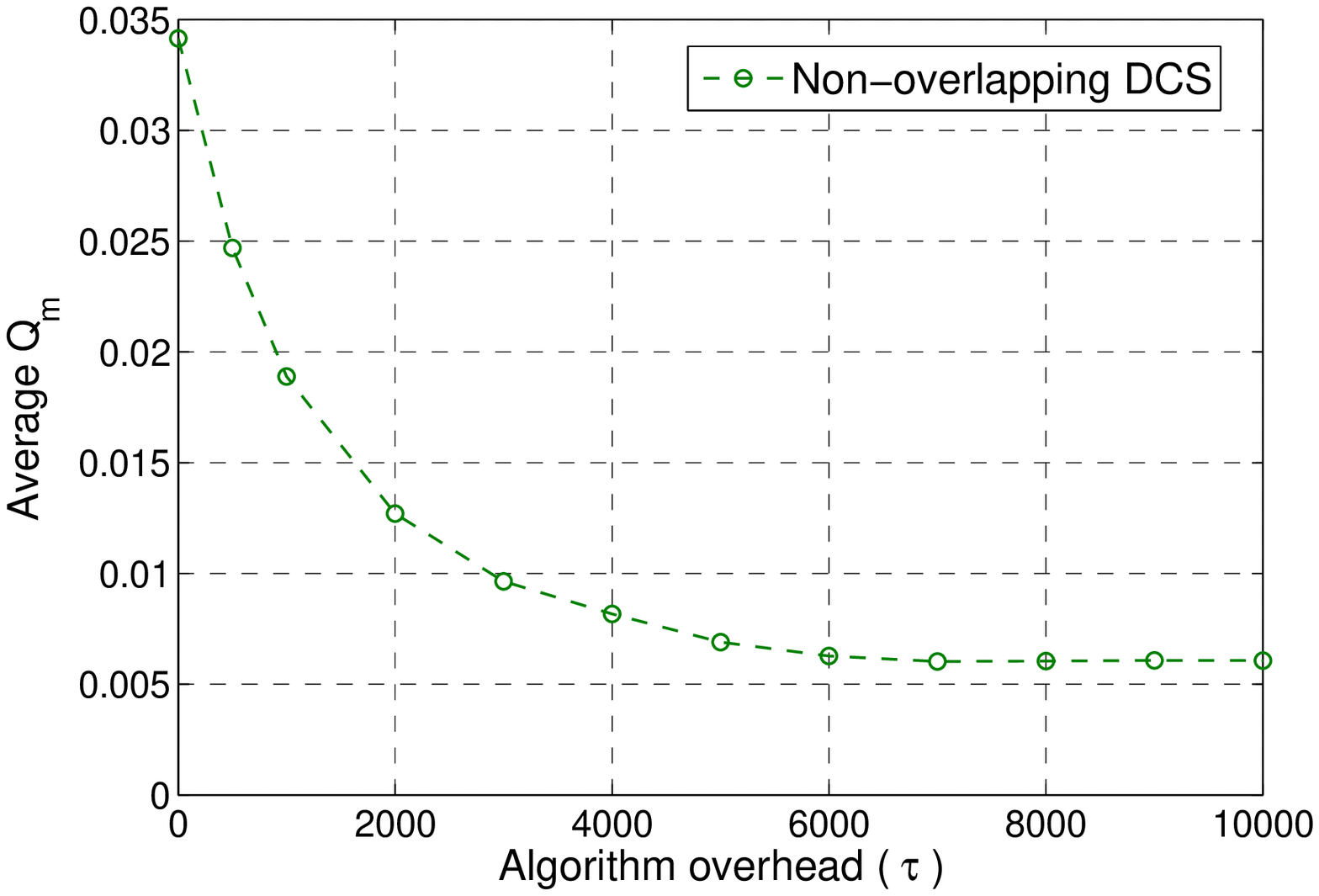}}
\caption{Network sensing performance as a function of the algorithm overhead with network size $N=30$, power constraint $P_{SU}=100$mW and bandwidth constraint $\theta_{SU}= 10$ in both criteria. The false alarm constraint in the $Q_m/Q_f$ criterion is $\alpha = 0.1$}
\label{PerformanceO} %% label for entire figure
\end{figure}

In Fig.~\ref{PerformanceO}, we show the network sensing performance as a function of the maximum overhead for each algorithm in each criterion for networks with $N=30$ SUs. For both the proposed algorithms, we can see that the network sensing performance improves fast as the SUs begin to exchange information, and then converges steadily to a final value. For the overlapping algorithm, $90\%$ improvement of the network sensing performance is obtained within the first $300\tau$ bits, while for the non-overlapping algorithm, it takes $4000\tau$ bits to achieve the same percentage.

\begin{figure}[!t]
\centering
\includegraphics[width=2.8in]{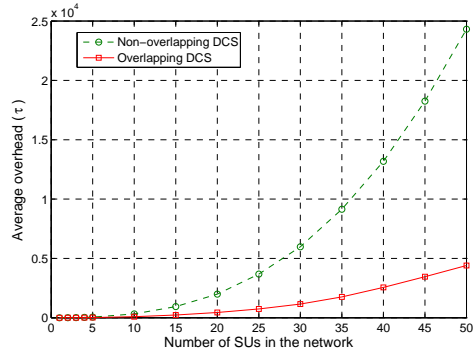}
\caption{Average overhead as a function of the number of SUs $N$ with power constraint $P_{SU} = 100$mW and bandwidth constraint $\theta_{SU} = 10$.} \label{overhead_N}
\end{figure}

In Fig.~\ref{overhead_N}, we show the overhead as a function of the number of SUs $N$ for each algorithm. In this figure, we can see that the overhead of the overlapping algorithm is only about $20\%$ of the non-overlapping algorithm, though the overlapping algorithm outperforms the non-overlapping algorithm in terms of network sensing performance as previously shown. In the non-overlapping algorithm, the basic operation involves the complete information of two coalitions, and the overhead is produced to check the feasibility of merge operations, as seen in (\ref{try}), even if the operation is not feasible. In the overlapping DCS, the feasibility of switch operation can be locally checked by the SU performing it, and no overhead is produced if the operation is not feasible. Therefore, the overlapping DCS needs less overhead than the non-overlapping DCS.

\begin{figure}[!t]
\centering
\includegraphics[width=2.8in]{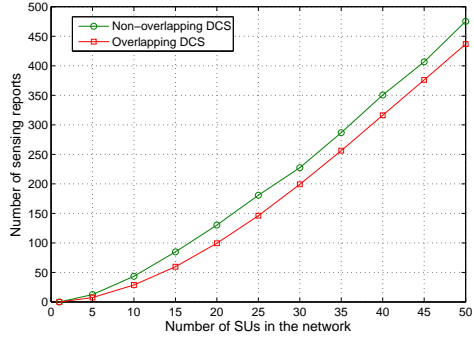}
\caption{Number of sensing reports as a function of the number of SUs $N$ with power constraint $P_{SU} = 100$mW and bandwidth constraint $\theta_{SU} = 10$.} \label{complexity_N}
\end{figure}

In Fig.~\ref{complexity_N}, we show the number of sensing reports as a function of the number of SUs $N$ for each algorithm. We see that the overlapping algorithm only needs $90\%$ sensing reports compared to the non-overlapping algorithm, which implies a lower system complexity for the cooperative sensing process. In the non-overlapping approach, an SU joins only one coalition, but it must report to all the members in this coalition. In the overlapping approach, although an SU may join multiple coalitions, it only needs to report to one SU for each coalition it joins, i.e., report to $i \in \mathcal{N}$ for joining coalition $\mathcal{R}_i$. Therefore, the overlapping approach, in which some SUs join multiple coalitions, does not necessarily imply an increase in system complexity.

%%%%%%%%%%%%%%%%%%%%%%
\section{Conclusions}%
%%%%%%%%%%%%%%%%%%%%%%

In this paper, we have proposed a novel approach for distributed cooperative sensing (DCS) with strict power and bandwidth constraints, in which the secondary users can form overlapping coalitions to optimize their spectrum sensing performance. In each coalition, a particular secondary combines the local sensing data from other coalition members and make a cooperative decision. The proposed algorithm is proved to converge to a stable outcome within finite iterations. Simulation results show that the proposed overlapping algorithm yields significant performance improvements, decreasing the total error probability up to $25\%$ in the $Q_m+Q_f$ criterion, the missed detection probability up to $20\%$ in the $Q_m/Q_f$ criterion, the overhead up to $80\%$, and the total report number up to $10\%$, compared with the state-of-the-art non-overlapping algorithm.

\appendices

\section{Optimal Threshold in the $Q_m+Q_f$ Criterion}

By substituting (\ref{Pm}), (\ref{Pf}) and $k_i = |\mathcal{R}_i|$ into (\ref{Qm}) and (\ref{Qf}), and then substituting (\ref{Qm}), (\ref{Qf}) and $\mathbf{\Lambda}(\mathcal{R}_i) = (\lambda, \lambda, \ldots, \lambda)_{1\times |\mathcal{R}_i|}$ into (\ref{utilityOrigin}), the utility function of the $Q_m+Q_f$ criterion is given as:
\begin{equation} \label{utilityA}
U(\mathcal{R}_i) =
1 - \min\limits_{\lambda} \left\{\left[\mathcal{Q}\left( \left( \lambda-1\right)\sqrt{N_s}\right) \right]^{|\mathcal{R}_i|} - \left[\mathcal{Q}\left( \left( \frac{\lambda}{1+\gamma}-1\right)\sqrt{N_s}\right) \right]^{|\mathcal{R}_i|} \right\}.
\end{equation}
The optimal $\lambda_a$ is the zero point of the first first-order derivative, and thus, it satisfies:
\begin{equation}
\frac{\partial}{\partial \lambda_a} \left[\mathcal{Q}\left( \left( \frac{\lambda_a}{1+\gamma}-1\right)\sqrt{N_s}\right) \right]^{|\mathcal{R}_i|} = \frac{\partial}{\partial \lambda_a} \left[\mathcal{Q}\left( \left( \lambda_a-1\right)\sqrt{N_s}\right) \right]^{|\mathcal{R}_i|}.
\end{equation}
By substituting $\mathcal{Q}'(x) = \exp{\left(-x^2/2\right)} / \sqrt{2 \pi}$, we have:
\begin{equation} \label{lambdaA}
\left[\frac{\mathcal{Q}\left( \left( \lambda_a-1\right)\sqrt{N_s}\right)}{\mathcal{Q}\left( \left( \frac{\lambda_a}{1+\gamma}-1\right)\sqrt{N_s}\right)}\right]^{|\mathcal{R}_i| - 1} - \frac{1}{1+\gamma} \exp{\left\{ \frac{N_s}{2} \left[ \left(\lambda_a-1\right)^2 - \left( \frac{\lambda_a}{1+\gamma} -1 \right)^2\right]\right\}} =0.
\end{equation}
Using (\ref{lambdaA}), the optimal threshold of the $Q_m+Q_f$ criterion can be evaluated numerically, denoted by $\lambda_a(|\mathcal{R}_i|)$. Note that the solution is only decided by the coalition size $|\mathcal{R}_i|$.

By substituting $\lambda = \lambda_a(|\mathcal{R}_i|)$ into (\ref{utilityA}), the utility function in (\ref{utilityOrigin}) for the $Q_m+Q_f$ criterion is also only determined by the coalition size $|\mathcal{R}_i|$, given by:
\begin{equation}
f_a(|\mathcal{R}_i|) =
1 - \left[\mathcal{Q}\left( \left( \lambda_a(|\mathcal{R}_i|)-1\right)\sqrt{N_s}\right) \right]^{|\mathcal{R}_i|} + \left[\mathcal{Q}\left( \left( \frac{\lambda_a(|\mathcal{R}_i|)}{1+\gamma}-1\right)\sqrt{N_s}\right) \right]^{|\mathcal{R}_i|}.
\end{equation}

\section{Optimal Threshold in the $Q_m/Q_f$ Criterion}

By substituting (\ref{Pm}), (\ref{Pf}) and $k_i = |\mathcal{R}_i|$ into (\ref{Qm}) and (\ref{Qf}), and then substituting (\ref{Qm}), (\ref{Qf}) and $\mathbf{\Lambda}(\mathcal{R}_i) = (\lambda, \lambda, \ldots, \lambda)_{1\times |\mathcal{R}_i|}$ into (\ref{utilityOrigin}), the utility function of the $Q_m/Q_f$ criterion is given as:
\begin{subequations}
\begin{align}
U(\mathcal{R}_i) &~ = \max\limits_{\lambda} \left[\mathcal{Q}\left( \left( \frac{\lambda}{1+\gamma}-1\right)\sqrt{N_s}\right) \right]^{|\mathcal{R}_i|} \\
s.t. &~ \left[\mathcal{Q}\left( \left( \lambda-1\right)\sqrt{N_s}\right) \right]^{|\mathcal{R}_i|} \le \alpha.
\end{align}
\end{subequations}
Note that $\mathcal{Q}(x)$ is a decreasing function with its value between $0$ and $1$. We can solve the constraint inequality as:
\begin{equation}
\lambda \ge \lambda_{min} = 1 + \frac{ \mathcal{Q}^{-1} \left(\alpha^{1/|\mathcal{R}_i|}\right) }{\sqrt{N_s}}.
\end{equation}
Also, since $\mathcal{Q}(x)$ is a decreasing function, the optimal threshold $\lambda_b$ is the minimal value $\lambda_{min}$, given by:
\begin{equation}
\lambda_b(|\mathcal{R}_i|) = 1 + \frac{ \mathcal{Q}^{-1} \left(\alpha^{1/|\mathcal{R}_i|}\right)}{\sqrt{N_s}}.
\end{equation}

By substituting $\lambda = \lambda_b(|\mathcal{R}_i|)$ into the objective function, we have the utility function in (\ref{utilityOrigin}) for the $Q_m/Q_f$ criterion:
\begin{equation}
f_b(|\mathcal{R}_i|) = \left[\mathcal{Q}\left( \frac{1}{1+\gamma} \left( \mathcal{Q}^{-1}\left( \alpha ^{1/|\mathcal{R}_i|}\right) - \gamma \sqrt{N_s} \right) \right) \right]^{|\mathcal{R}_i|}.
\end{equation}

\def\baselinestretch{1.1}

\end{document}